\documentclass[reprint, superscriptaddress, amsmath, amssymb, aps, pra]{revtex4-2}
\usepackage{graphicx}
\usepackage{dcolumn}%
\usepackage{xcolor}
\usepackage[absolute]{textpos}
\definecolor{darkblue}{rgb}{0,0.3,0.7}
\usepackage{hyperref, comment}
\hypersetup{
colorlinks=true,
linkcolor=darkblue,
filecolor=blue,
citecolor=darkblue,  
urlcolor=darkblue,
}

\definecolor{ASUblue}{RGB}{0,163,224}

\DeclareMathOperator{\Tr}{Tr}

\begin{document}

\preprint{APS/123-QED}

\title{
Quantum nonlocal modulation cancellation with distributed clocks
}
\author{Stephen D. Chapman}
 \affiliation{Elmore Family School of Electrical and Computer Engineering and Purdue Quantum Science and Engineering Institute, Purdue University, West Lafayette, Indiana 47907, USA}
\author{Suparna Seshadri}
 \affiliation{Elmore Family School of Electrical and Computer Engineering and Purdue Quantum Science and Engineering Institute, Purdue University, West Lafayette, Indiana 47907, USA}
\author{Joseph~M. Lukens}
\affiliation{Research Technology Office and Quantum Collaborative, Arizona State University, Tempe, Arizona 85287, USA}
\affiliation{Quantum Information Science Section, Computational Sciences and Engineering Division, Oak Ridge National Laboratory, Oak Ridge, Tennessee 37831, USA}
\author{Nicholas~A. Peters}
\affiliation{Quantum Information Science Section, Computational Sciences and Engineering Division, Oak Ridge National Laboratory, Oak Ridge, Tennessee 37831, USA}
\author{Jason~D. McKinney}
\author{Andrew~M. Weiner}
 \affiliation{Elmore Family School of Electrical and Computer Engineering and Purdue Quantum Science and Engineering Institute, Purdue University, West Lafayette, Indiana 47907, USA}
 \author{Hsuan-Hao Lu}
  \email{luh2@ornl.gov} 
\affiliation{Quantum Information Science Section, Computational Sciences and Engineering Division, Oak Ridge National Laboratory, Oak Ridge, Tennessee 37831, USA}

\date{\today}

\begin{textblock}{13.3}(1.4,15)
\noindent\fontsize{7}{7}\selectfont \textcolor{black!30}{This manuscript has been co-authored by UT-Battelle, LLC, under contract DE-AC05-00OR22725 with the US Department of Energy (DOE). The US government retains and the publisher, by accepting the article for publication, acknowledges that the US government retains a nonexclusive, paid-up, irrevocable, worldwide license to publish or reproduce the published form of this manuscript, or allow others to do so, for US government purposes. DOE will provide public access to these results of federally sponsored research in accordance with the DOE Public Access Plan (http://energy.gov/downloads/doe-public-access-plan).}
\end{textblock}

\begin{abstract}
We demonstrate nonlocal modulation of entangled photons with truly distributed RF clocks. Leveraging a custom radio-over-fiber (RFoF) system characterized via classical spectral interference, we validate its effectiveness for quantum networking by multiplexing the RFoF clock with one photon from a frequency-bin-entangled pair and distributing the coexisting quantum-classical signals over fiber. Phase modulation of the two photons reveals nonlocal correlations in excellent agreement with theory: in-phase modulation produces additional sidebands in the joint spectral intensity, while out-of-phase modulation is nonlocally canceled. Our simple, feedback-free design attains sub-picosecond synchronization---namely, drift less than $\sim$0.5~ps in a 5.5~km fiber over 30~min (fractionally only $\sim$2$\times$10$^{-8}$ of the total fiber delay)---and should facilitate frequency-encoded quantum networking protocols such as high-dimensional quantum key distribution and entanglement swapping, unlocking frequency-bin qubits for practical quantum communications in deployed metropolitan-scale networks.
\end{abstract}

\maketitle

\section{Introduction}
Nonlocality is a fundamental and counterintuitive feature of quantum theory~\cite{Einstein1935, Bell1964, Maudlin2019}. Observed in Bell inequality tests~\cite{Aspect1981, Aspect1982a, Aspect1982b, Hensen2015, Giustina2015, Shalm2015} and leveraged in quantum information processing tasks such as quantum key distribution (QKD)~\cite{Ekert1991, Gisin2002} and teleportation~\cite{Bennett1993, Bouwmeester1997}, the puzzling fact that a particle's state can depend on measurements performed on an entangled partner arbitrarily far away has, paradoxically, proven central to both the strangeness and utility of quantum mechanics. Yet although a fundamentally \emph{quantum} phenomenon, the measurement and utilization of nonlocality place stringent requirements on \emph{classical} resources as well, particularly in terms of synchronization between distributed sites---at the very least for timestamping detection events, and perhaps additionally in coordinating operations performed on entangled particles at multiple receivers.

The precision required for such temporal coordination depends on both the physical encoding and the application. For example, in Bell tests of polarization-entangled photons, the target timescales for event counting are related to detector jitter (typically on the order of 100~ps~\cite{Hadfield2009}, though 3~ps has been demonstrated~\cite{Korzh2020}), while active basis choices must be faster than the time of flight from source to receiver to close the locality loophole~\cite{Aspect1982b, Giustina2015, Shalm2015}. On the other hand, the intrinsically temporal nature of time-frequency encodings imply synchronization requirements related to the photonic wavepackets themselves. Fast pulse-by-pulse switches are required to convert passive but probabilistic time-bin interferometers~\cite{Islam2017a, Islam2017b} to theoretically unit success probability~\cite{Wang2015, Wang2018}, while frequency-bin-encoded states~\cite{Lu2023c}---the focus of this work---can be effectively processed by electro-optic phase modulators (EOPMs) driven by radio frequency (RF) fields at the bin spacing~\cite{Lukens2017, Kues2017,  Lu2018a, Imany2018, Lu2018b}. Nonlocal phenomena such as modulation cancellation can be observed in this encoding as well~\cite{Harris2008}, which relies on the fact that the spectral correlations of frequency-entangled photons depend on the cumulative temporal phase modulation experienced by both, independent of their spatial separation~\footnote{Note that past demonstrations of quantum nonlocal cancellation, including both dispersion~\cite{Franson1992, Baek2009, Lee2014} and modulation cancellation~\cite{Harris2008, Sensarn2009, Zhou2023}, have so far not attempted to close the locality loophole (encountered in Bell tests) by manipulating photons with spacelike separation. Yet nothing fundamentally prevents such spacelike separation from being realized.}. Consequently, coherent out-of-phase modulation on each photon can cancel such that the joint spectral correlations appear as though no modulation occurred~\cite{Sensarn2009}. In order to observe this effect at typical bin or filter widths $\gtrsim$10~GHz, such nonlocal modulation requires distributed RF signals with jitters much smaller than an RF period (i.e., $\ll$100~ps).

\begin{figure*}[t!]
\includegraphics[width=\textwidth]{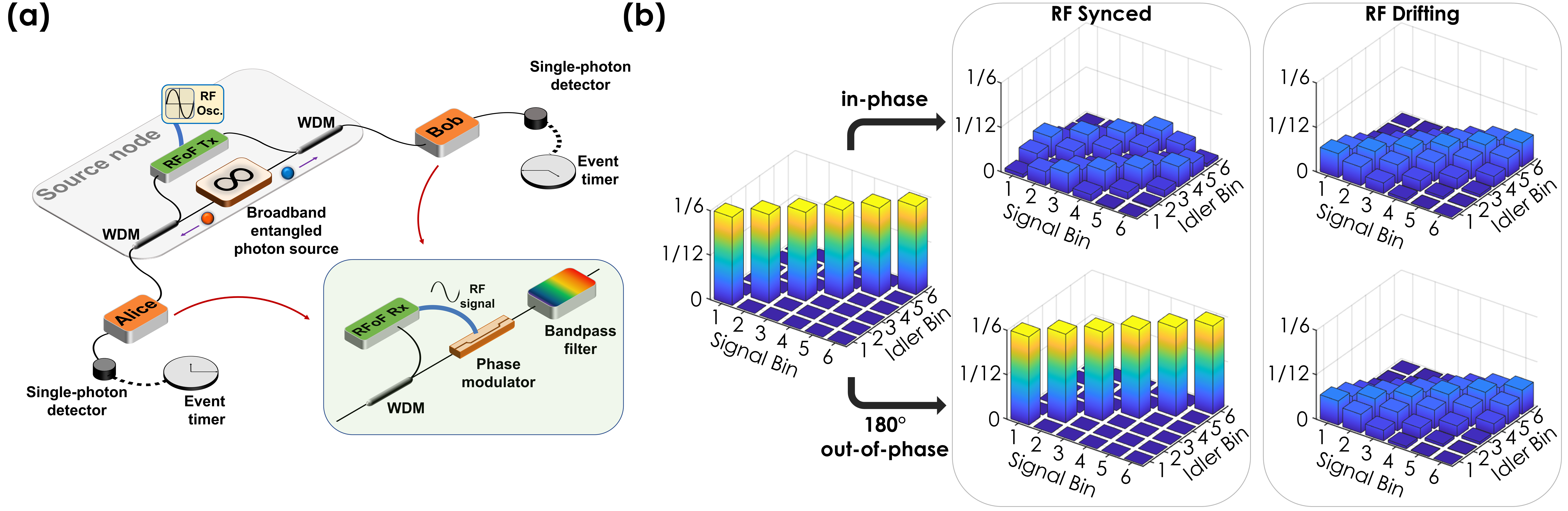}
\caption{(a) Conceptual illustration of temporal modulation of frequency-entangled photons with distributed clocks. (b) Simulated JSI of frequency-bin-entangled states subjected to in-phase and out-of-phase modulation, with either synchronized (left) or uniformly drifting (right) RF signals. WDM, wavelength-division multiplexer; RFoF Rx/Tx, RF-over-fiber receiver/transmitter; Osc.,  oscillator.}
\label{concept}
\end{figure*}

Although there exist several demonstrations of high-speed modulation of spatially separated photons~\cite{Sensarn2009, Olislager2010, Sabattoli2022,Seshadri2022, Lu2023b, Borghi2023, Clementi2023} all experiments so far have sidestepped the challenge of timing synchronization entirely, by using a single RF oscillator whose signals are delivered directly to each EOPM. Due to RF cable losses, such configurations are limited in practice to local scenarios in a single laboratory, and are therefore incompatible with the extension of frequency-bin encoding to deployed quantum networks. Figure~\ref{concept}(a) illustrates such a vision. At the source node, frequency-bin-entangled signal and idler photons are generated and launched into deployed fiber toward two users, Alice and Bob. Each user is equipped with an EOPM and bandpass filters to realize specific frequency projections. For this purpose, Alice and Bob must drive their individual EOPMs with RF signals that remain phase-locked throughout the integration time of their measurements. Given their physical separation, one natural solution would be to leverage an RF-over-fiber (RFoF) system coexisting with the quantum signal of interest via wavelength-division multiplexing to distribute synchronized RF signals from the source node to the users over fiber.

In this work, we solve these challenges and demonstrate nonlocal modulation of entangled photons with truly distributed RF clocks. Constructing a custom RFoF system, we distribute a 19~GHz tone through intensity modulation and direct detection, the performance of which we characterize through sideband measurements in an ``identity gate''---two out-of-phase EOPMs separated by various lengths of optical fiber---inferring timing variations less than 0.5~ps over 30~min in spools up to 5.5~km. By multiplexing the RFoF clock with one photon of an entangled pair, transmitting both through a fiber channel, and then separating the classical signal upon receipt to modulate the entangled photon, we successfully demonstrate nonlocal modulation cancellation~\cite{Harris2008} over 200~m in optical fiber. Not only does our experiment realize modulation cancellation in a configuration compatible with nonlocality but our scalable RFoF technique removes what has arguably proven the most conspicuous barrier to distributed frequency-bin quantum information, opening the door for quantum frequency processing in future quantum networks.


\section{Background}
\label{sec:background}
Under the overall umbrella of microwave photonics~\cite{Yao2009, Urick2015}, a variety of techniques for distributing RF clocks over optical fiber have been demonstrated. Yet existing techniques generally fall into two extremes: 
(i) inexpensive and scalable components, but with relatively high jitter ($\gtrsim$1~ps); or (ii) ultralow (sub-femtosecond) jitters, but with complex and expensive optical control systems.

An example of the former, White Rabbit~\cite{Lipinski2011} is an open-source extension of the Precision Time Protocol~\cite{IEEE2019} that has successfully distributed 10~MHz clocks with jitter as low as 1.1 ps (0.8 ps) root-mean-square, integrated from 1--30~Hz (10$^2$--10$^5$~Hz) over 10~km of fiber~\cite{Rizzi2018}. Although this performance is more than sufficient for time tagging in quantum networks~\cite{Alshowkan2022b, Burenkov2023}, employing such clocks to facilitate the distribution or synchronization of tens of GHz RF signals remains a challenge. Indeed, an attempt to synchronize two 19~GHz RF signals (from two independent oscillators) via White Rabbit proved unsatisfactory, as detailed in Sec.~\ref{sec:classical2} below. On the other hand, advanced optical techniques combining features such as active phase stabilization, low-linewidth lasers, or even octave-spanning frequency combs can do much better~\cite{Foreman2007}, with timing jitters in the attosecond regime demonstrated over km-scale distances~\cite{Xin2017}.

Missing from these two extremes are middle-ground options that can combine the simplicity of (i) with sub-picosecond jitters, for which the best demonstrations of (ii) overperform. In developing our solution, we approach the problem guided by three foci: 
\begin{enumerate}
\item \emph{High-frequency operation.} We ultimately seek clock signals $\sim$20~GHz for modulating frequency bins that can be readily resolved by commercial wavelength-selective switches (WSSs). Rather than distributing a lower frequency (e.g., 10~MHz) and multiplying or locking a second oscillator to reach the GHz regime, we propose to distribute the frequency of interest directly. This simplifies the microwave system, reduces cost, and intuitively suppresses jitter due to the fast rise times of GHz optical signals.
\item \emph{Relative time synchronization.} Our goal is not to synchronize remote sites to a global time, but rather coordinate modulation on received photons. 
Thus, if the delay experienced by the quantum signal drifts due to environmental fluctuations, it is actually advantageous for the RFoF phase to drift commensurately, which can be facilitated by multiplexing quantum and classical signals in nearby bands in the same fiber.
\item \emph{Sideband-based characterization.} Typical metrics for characterizing RFoF clock distribution include Allan deviation and integrated jitter. However, in the targeted application of frequency-bin processing, fidelity is most readily observed in the frequency domain, particularly the suppression of unwanted sidebands in the frequency-bin state. As shown quantitatively in Appendix~\ref{AppSim}, sideband suppression is still related to RF jitter. Thus, we develop an equivalent characterization approach based on optical modulation: two EOPMs separated by optical fiber, with the first EOPM driven by the reference clock and the second by the distributed clock.
\end{enumerate}

\begin{figure*}[t!]
\includegraphics[width=\textwidth]{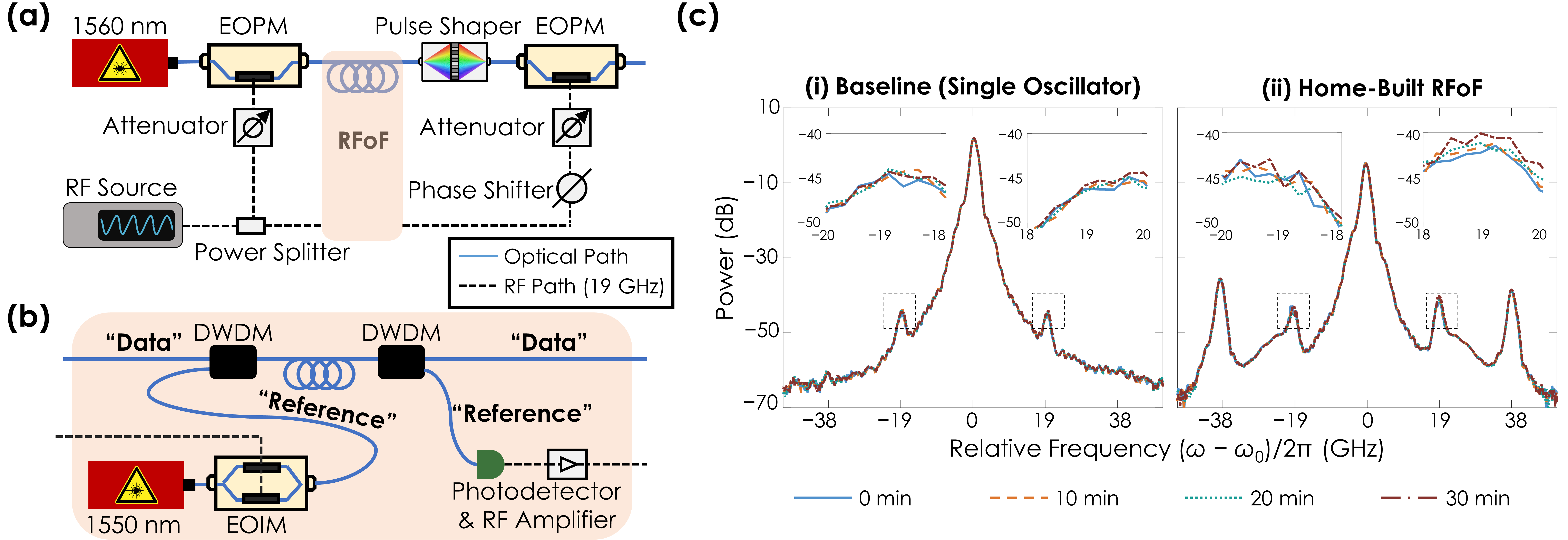}
\caption{(a) Experimental setup for classical modulation cancellation. This configuration captures the relative drifts between the RF signals, which are reflected in the optical spectrum at the output via spectral interference. (b) Diagram of our home-built RFoF system. (c) Measured optical spectra when (i) RF signals from a single oscillator are split to directly drive both EOPMs and (ii) our home-built RFoF system is used to distribute the RF signals. The figure insets provide a zoomed-in view of the optical spectra near the first-order sidebands. DWDM, dense wavelength-division multiplexer; EOPM, electro-optic phase modulator; EOIM, electro-optic intensity modulator.}
\label{classic}
\end{figure*}

\section{Classical Tests}
Figure~\ref{classic}(a) depicts the basic approach used to benchmark our RFoF clock distribution system via cascaded spectral interference in the optical domain. In this setup, a continuous-wave (CW) laser operated at $\sim$1560~nm (termed ``data channel'') passes through two EOPMs in series, accumulating sinusoidal temporal phase modulation. These EOPMs are connected via a specific length of optical fiber and each is driven by RF sinewaves that we intend to synchronize. When both EOPMs apply identical but $180^{\circ}$ out-of-phase modulation to the transmitted light, the frequency sidebands generated by the first EOPM destructively interfere at the output of the second EOPM, effectively returning the original monochromatic spectrum. By monitoring the output spectrum, we can assess the stability of the two clocks. Notably, any deterioration in the contrast between the original spectral line and the sidebands indicates a drift in the relative phase between the two RF signals.

\subsection{Local Benchmark}
\label{sec:classical1}
In the initial test, we drive both EOPMs with RF signals derived from the same oscillator (Agilent E8257D), which ideally should introduce minimal temporal drift and establish a baseline for subsequent RFoF tests.  
In all tests in this section, we drive both EOPMs at 19~GHz with a modulation depth of 1.42 radians---a setting inspired by the probabilistic Hadamard operation which produces an output whose first-order sidebands are equal to the carrier~\cite{Lu2020b}. By carefully adjusting an RF variable attenuator and a phase shifter, we aim to minimize frequency sidebands at the output, ensuring that the temporal modulation on both EOPMs are perfectly aligned yet $180^{\circ}$ out of phase. In the ``local'' test configuration, the 19~GHz source is split via an RF power splitter and distributed directly to the first and second EOPMs over coaxial cable. This configuration achieves a sideband suppression of $-$48.08~dBc~\footnote{In this work, we define the amount of sideband suppression, or contrast, as the difference between the optical power in the original frequency mode and the power in highest first-order sideband.}, as shown in Fig.~\ref{classic}(c), with first-order sidebands exhibiting less than 2~dB fluctuation over 30 min. These fluctuations may stem from variables such as optical polarization drift, slight backlash in the RF attenuator, and measurement uncertainty in the optical spectrum analyzer (OSA).

\subsection{Nonlocal System Design}
\label{sec:classical2}
The experimental setup used in the baseline test presents challenges when the EOPMs are physically separated across different quantum nodes, primarily due to the significant loss introduced by coaxial cables, which are impractical for directly transmitting high-frequency RF signals over long distances. For instance, a typical $1$~m SMA coaxial cable can incur up to $\sim$1.2~dB loss in the $K_u$-band (12--18 GHz)~\cite{Pasternack}, and is significantly more expensive than single-mode fibers of the same length. In general, one can approach the clock distribution problem in two basic ways, by transmitting via RFoF either (i)~a low-frequency reference (e.g., 10~MHz) that is then used to lock a higher-frequency oscillator at the remote node or (ii)~the microwave frequency of interest (19~GHz in our case) for direct modulation of the quantum signal. 

To explore the feasibility of (i) with commercial options, we consider individual oscillators (Agilent E8257D) synchronized by a shared 10~MHz reference. We first consider a direct BNC coaxial connection ($<$3~m) between the two RF sources; while not scalable to large distances, this situation provides a straightforward benchmark 
for subsequent RFoF solutions.
Experimentally, we observe sideband cancellation down to $-42.90$~dBc, but it degrades to $-17.36$~dBc after just 5~min (corresponding to 3~ps of drift; Appendix \ref{AppSim}), even though the oscillators are only a few meters apart. Accordingly, to support any quantum experiment requiring long integration times, some form of feedback loop will be necessary to compensate for such drifts. Proceeding to a true RFoF solution,
we introduce two White Rabbit (WR) modules---a WR leader and a WR follower---for distributing a 10~MHz reference to both oscillators. We position the WR leader near the first oscillator and the WR follower near the second, with each WR outputting a locked 10 MHz signal. These two WR nodes are connected through 2~m of single-mode fiber and synchronized using the high-accuracy Precision Time Protocol~\cite{IEEE2019}, allowing the oscillators to be separated by greater distances compared to the 10~MHz direct link discussed above. However, we observe that the optical power in the sidebands fluctuates more rapidly in this configuration, even within the OSA's integration time ($\sim$1~s), preventing any effective cancellation. 

While, these 10~MHz tests are certainly far from exhaustive; it is quite likely that more advanced 10~MHz RFoF systems or tailored frequency multiplier circuits could enable more reliable synchronization between the two nodes than observed either with a coaxial or WR-based connection. However, our results do confirm the challenges associated with simple 10~MHz-based locking with inexpensive off-the-shelf components: stability at 10~MHz does not readily translate to stability at 19~GHz. This observation is consistent with the findings in the Appendix C of Ref.~\cite{Chapman2023}.
 
Accordingly, we introduce a custom RFoF technique to share high-frequency RF signals from a single oscillator between two nodes, circumventing the limitations previously discussed. Figure~\ref{classic}(b) depicts our home-built RFoF system, which utilizes intensity modulation and direct detection. 
In this system, the 19~GHz RF signal from the oscillator is split into two paths---one feeds the first EOPM, and the other drives an electro-optic intensity modulator (EOIM), which is biased at the quadrature point (i.e., 50\% transmission). A CW laser operating at $\sim$1550 nm (termed ``reference channel'')  passes through the EOIM, transferring the sinusoidal drive voltage onto optical intensity. A dense wavelength-division multiplexer (DWDM) combines this intensity-modulated reference with the phase-modulated ``data channel'' into a single fiber, facilitating the transmission of the multiplexed channels to the second node.

At the second node, another DWDM demultiplexes these two channels into separate fibers. The data channel undergoes dispersion compensation via a pulse shaper before entering the EOPM. The reference channel, meanwhile, is directed to a high-speed photodetector (BPR-23-M; Optilab) that extracts the 19~GHz sinusoidal RF signal, which is then used to drive the second EOPM. We finely adjust the RF power driving the EOIM to maximize the power in the 19~GHz output tone. Additionally, an electrical spectrum analyzer is used [not shown in Fig.~\ref{classic}(b)] to continuously monitor that the EOIM is accurately biased at the quadrature point.

We first test this configuration with $\sim$2~m of fiber connecting the two DWDMs. Figure~\ref{classic}(c) presents the results, which demonstrate an initial suppression of $-38.15$~dBc with less than 3~dB drift over the following 30 min, indicating successful RF synchronization between the two EOPMs. Subsequently, we extend the separation between nodes by inserting fiber spools of varying lengths between the DWDMs. For each length, we program the pulse shaper to apply an optimized quadratic spectral phase to the data channel, compensating for group velocity dispersion. Stable sideband cancellation is achieved in all four scenarios tested (200~m, 700~m, 1.1~km, and 5.5~km of single-mode fiber), with detailed results plotted in Appendix~\ref{plots}. Since the data and reference channels share the same fiber in our design, thermal fluctuations in the common path are effectively neutralized, underscoring the capability of our RFoF system for long-distance RF signal distribution without the need for any external feedback.

Finally, we have developed a simulation tool that translates the degree of sideband cancellation into an estimation of how much the two RF signals drift relative to each other. As detailed in Appendix~\ref{AppSim}, this tool enables us to input a timing offset between two EOPMs and calculates the corresponding sideband amplitudes, and vice versa. Utilizing data from the 5.5~km fiber spool test, the simulation estimates a drift $<$0.5~ps, or less than 1\% of the RF period. 

\section{Nonlocal Modulation Cancellation}

\begin{figure}[t!]
\includegraphics[width = \columnwidth]{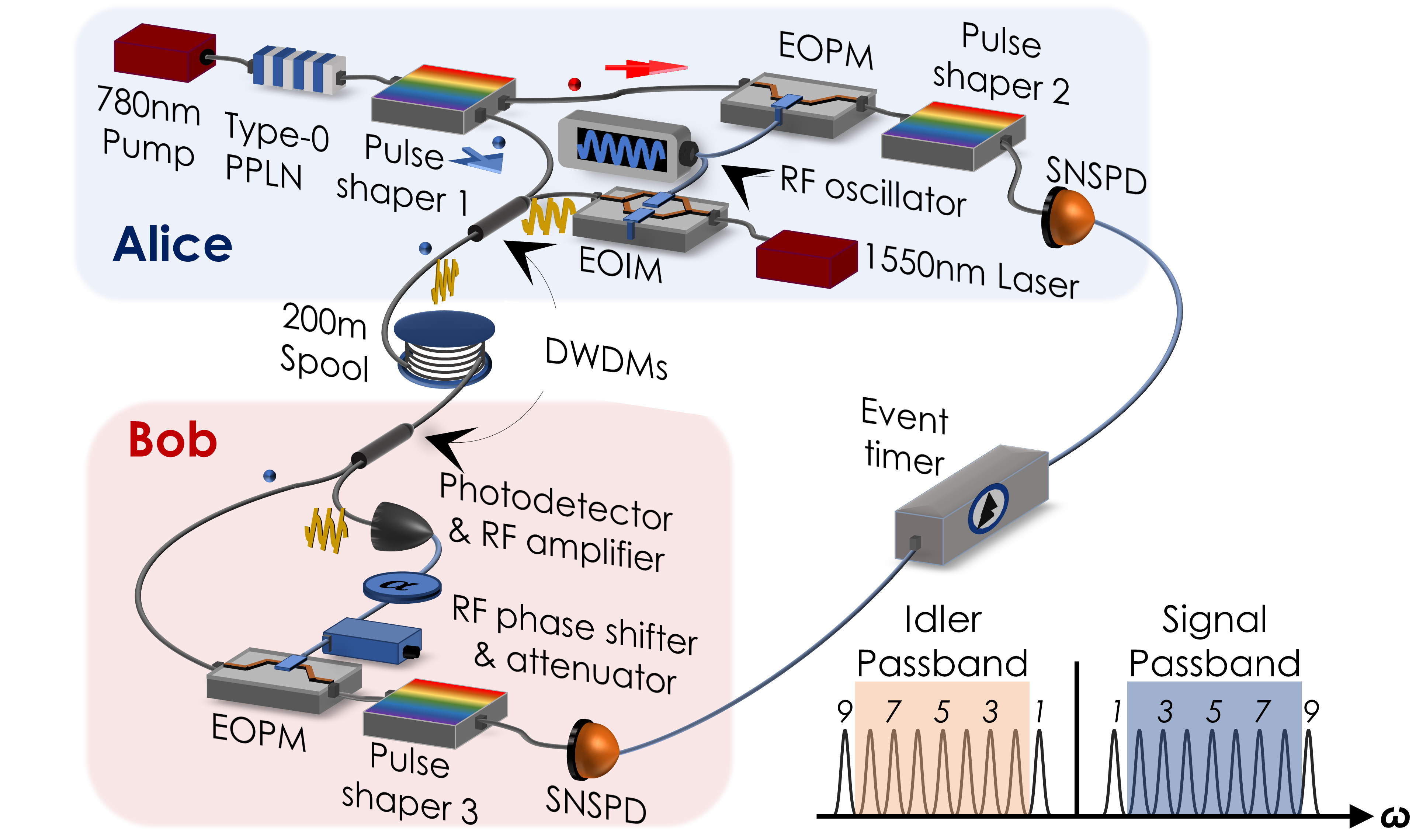}
\caption{Experimental setup for nonlocal modulation cancellation. Entangled photons and classical RFoF clock are represented by spheres and sinewaves, respectively.  The inset displays the passbands programmed on Pulse Shaper 1, and the spectral filters (or ``bins'') scanned on Pulse Shapers 2 and 3 for JSI measurements. PPLN, periodically poled lithium niobate waveguide; SNSPD, superconducting nanowire single-photon detector.}
\label{Qsetup}
\end{figure}

The long-term stability and sub-picosecond jitter demonstrated in our classical tests suggest that our RFoF system could be extremely valuable in quantum experiments requiring the synchronization of high-frequency RF signals across spatially separated nodes. Our design can be seamlessly integrated into these experiments, ensuring a consistent phase relationship for operations performed \emph{nonlocally} on photonic states. We demonstrate these capabilities by revisiting a foundational experiment for time-energy-entangled photons---the nonlocal modulation cancellation effect~\cite{Harris2008, Sensarn2009}. In this experiment, frequency-entangled signal and idler photons are subjected to temporal phase modulation in two separate EOPMs, which are now synchronized through our RFoF system. Figure~\ref{concept}(b) illustrates this concept by simulating the joint spectral intensity (JSI) of a high-dimensional frequency-entangled state when signal and idler photons are modulated by identical phase modulations that are either in phase or $180^{\circ}$ out of phase. In the ideal scenario, in-phase modulations will spread out the frequency correlations due to newly generated sidebands, while out-of-phase modulations will return the JSI to its original form, as their effects on the joint spectral correlation cancel each other out. Conversely, in the most extreme scenario where RF synchronization is unsuccessful and these modulations drift over multiple RF periods during the measurement, the JSIs in both scenarios will appear identical, as they represent an incoherent sum of JSIs at various timing offsets.

Figure~\ref{Qsetup} illustrates the experimental setup for nonlocal modulation cancellation. We utilize a CW laser (DL pro; Toptica) operating at 778.5 nm to pump a 2~cm-long periodically poled lithium niobate (PPLN; AdvR) ridge waveguide, generating broadband, frequency-entangled photons via type-0 spontaneous parametric down-conversion. The signal and idler photons are separated by frequency and routed to two output fibers using the first pulse shaper (WaveShaper 4000A; Coherent). Additionally, we program two 140~GHz-wide bandpass filters on both the signal and idler spectra to match the passbands of the subsequent DWDMs in the idler arm. The signal photons are directed to an EOPM for phase modulation at Alice's node, while the idler photons, multiplexed with the intensity-modulated reference laser from the RFoF system, are transmitted to Bob's node. Due to the fact our superconducting nanowire single-photon detectors (SNSPDs) are colocated, Alice and Bob are currently located in the same room and connected by a 200~meter fiber spool. However, our demonstration is fully capable of extending to truly physically separated nodes linked through deployed fibers.

\begin{figure}[t!]
\includegraphics[width = \columnwidth]{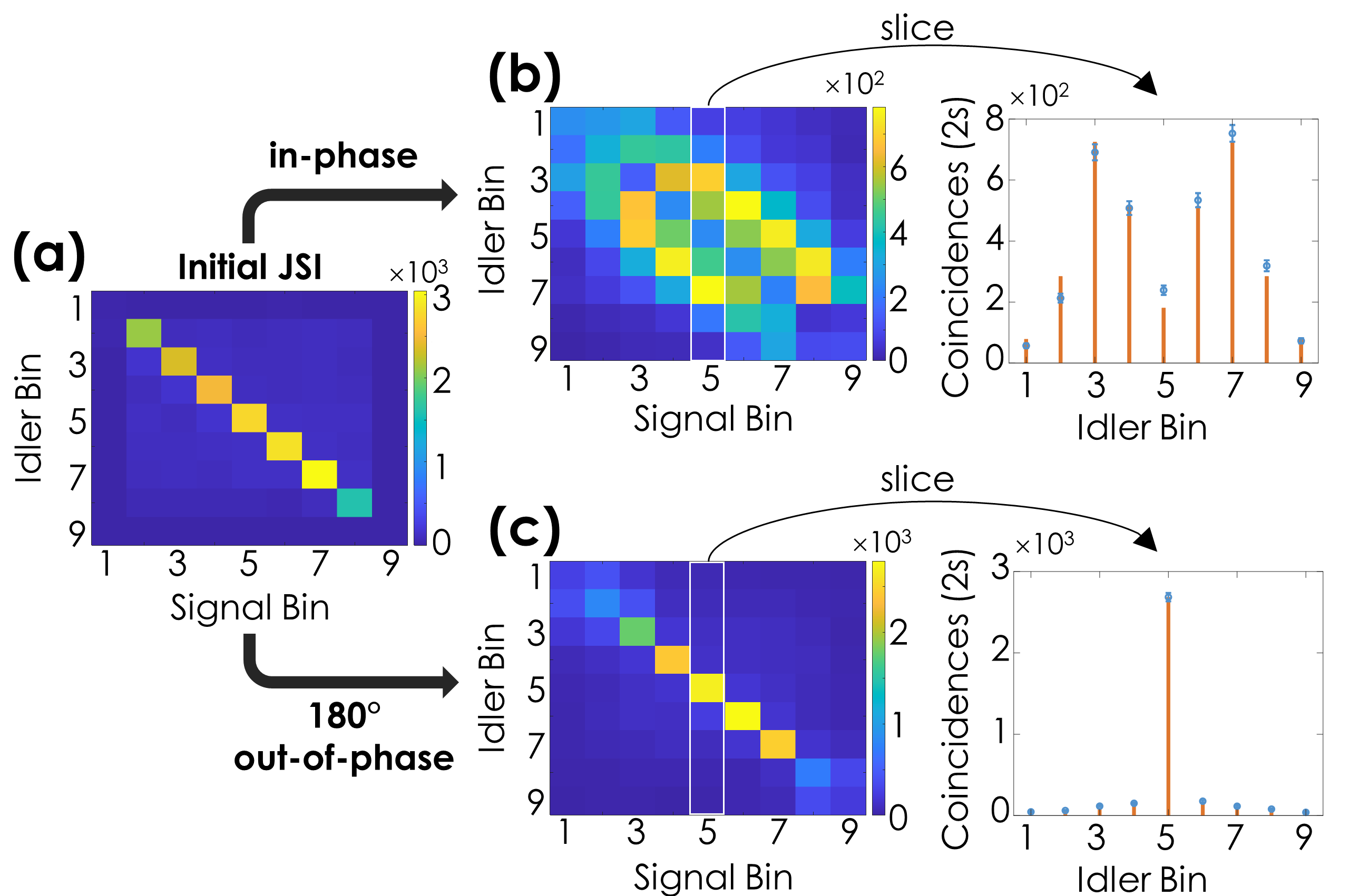}
\caption{Measured JSIs for three scenarios: (a) RF oscillator off and both EOPMs unmodulated, (b) EOPMs synchronously driven in phase, and (c) EOPMs synchronously driven $180^{\circ}$ out of phase. Coincidences are integrated over 2~s per bin pair (shown in blue), with error bars assuming Poissonian statistics. The red stem plots represent theoretical predictions, scaled and vertically offset to match the data points via linear least squares. The reference laser remains on for all three scenarios.}
\label{quantumJSI}
\end{figure}

At Bob's node, due to the contrast in brightness between the quantum state and the classical channel ($\sim$4 mW before the EOIM), we cascade two identical DWDMs to separate idler photons from the reference channel and further minimize the crosstalk noise infiltrating the quantum output. The idler photons are subsequently directed to an EOPM, which is driven by RF waveforms generated from the RFoF system. At the output of each EOPM, the phase-modulated photons are routed through another pulse shaper (WaveShaper 1000S; Coherent Corp.). We program and scan spectral filters on the pulse shapers to collect coincidences over a $9\times9$ frequency-bin grid, effectively conducting a JSI measurement. Each bin in this grid is 12~GHz wide and spaced by 19~GHz. Experimentally, we measure approximately $10^3$ s$^{-1}$ noise photons in each 12~GHz bin caused by the classical channel. Initially, we deactivate the RF oscillator, leaving both EOPMs unmodulated, and record the initial JSI of the biphotons. The results reveal a total of 7 frequency-bin pairs within the 140~GHz passband [Fig.~\ref{quantumJSI}(a)]. 

When the EOPMs in the signal and idler arms are driven synchronously, the phase modulation applied to individual photons cumulatively influences their biphoton wavefunction---a manifestation of time-energy entanglement~\cite{Harris2008, Sensarn2009} and likewise observable through JSI measurements. We again apply 19~GHz sinewave modulation at a modulation depth of $1.42$~rad to both EOPMs, and carefully adjust their relative timing through the RF phase shifter to explore two specific conditions: in-phase and $180^{\circ}$ out-of-phase modulation. In the in-phase scenario, the two phases add up on the entangled photon pair and, as expected, introduce newly generated sidebands that significantly alter the JSI [Fig.~\ref{quantumJSI}(b)]. 

On the other hand, when the two RF sinewaves are $180^{\circ}$ out of phase, they destructively interfere, canceling each other out as if the photon pairs were subjected to no modulation at all. As illustrated in Fig.~\ref{quantumJSI}(c), the measured JSI in this condition closely resembles the initial state [Fig.~\ref{quantumJSI}(a)]. The sidebands are significantly suppressed, resulting in coincidence counts that largely remain on the diagonal. Some discrepancies are observed in the bin pairs near the edge of the filter passband programmed on Pulse Shaper 1, as they receive fewer coincidences. Ideally, these bin pairs would capture frequency-shifted photons originating from outer frequency bins, such as signal and idler bins 1 and 9 shown in the inset of Fig.~\ref{Qsetup}. However, these bins are now blocked by the first pulse shaper as they lie outside of the passband. 

Assuming the spectral filter on one of the photons remains fixed, isolating only the central bin while scanning the filter on the other photon, the normalized coincidences should theoretically follow the first-order Bessel function $J^{2}_{n}(2\delta)$ for the in-phase scenario and $J^{2}_{n}(0)$ for the out-of-phase scenario~\cite{Harris2008, Sensarn2009}, where $n$ denotes the frequency bin located $19n$~GHz away from the central bin. To refine the model to account for experimental imperfections, we add the effects of filter crosstalk (due to finite pulse shaper resolution) into the theory function, scale it to reflect the biphoton flux, and vertically offset it to account for accidentals; we then fit this function to the experimental data via linear least squares. Here, the accidentals primarily stem from multipair emission rather than from the RFoF system: registered singles count rates from the entangled-photon source are more than 100 times greater than the crosstalk noise. In both cases in Fig.~\ref{quantumJSI}(b,c), we find that our experimental results closely match the model, demonstrating successful distribution of high-frequency clocks for nonlocal modulation of biphotons.

\section{Discussion}

Moving forward, we anticipate no obstacles to integrating our system into deployed networks. In such scenarios, signal and idler photons, along with the classical RFoF channel, will be transmitted to spatially separated nodes through deployed fiber. At each node, photons will ideally be detected by individual SNSPD modules (unavailable in this proof-of-concept demonstration), with precise time tagging of detection events enabled by the shared RFoF clock or perhaps parallel WR-based timing distribution~\cite{Alshowkan2022b, Burenkov2023}. Our design could be particularly beneficial for quantum networking protocols based on frequency-encoded photons, such as high-dimensional QKD, quantum teleportation, and entanglement swapping.

For example, in high-dimensional (two-basis) QKD~\cite{Sheridan2010}, pairs of frequency-entangled qudits can be generated and distributed to separated quantum nodes, each equipped with a frequency-bin processing unit like the quantum frequency processor (QFP)~\cite{Lu2018a, Lu2018b, Lu2023c}, which facilitates measurements of incoming photons in both the computational and discrete Fourier transform (DFT) bases. In such cases, the high-speed EOPMs in these individual QFPs must be synchronized precisely to ensure the error probability remains well below the asymptotic bound. In Appendix \ref{sec:appC}, we investigate the fidelity of DFT gates synthesized on the QFP circuit in the presence of additional RF delay and find that the degree of synchronization demonstrated by our RFoF system is capable of supporting the processing of frequency-bin qubits and high-dimensional qudits in metro-area networks.

\begin{acknowledgments}
We thank M. Alshowkan for helpful discussions regarding White Rabbit. This work was performed in part at Oak Ridge National Laboratory, operated by UT-Battelle for the U.S. Department of energy under contract no. DE-AC05-00OR22725. Funding was provided by the U.S. Department of Energy, Office of Science, Advanced Scientific Computing Research (Field Work Proposals ERKJ381, ERKJ353).
\end{acknowledgments}

%

\appendix

\begin{figure}[t!]\includegraphics[width=\columnwidth]{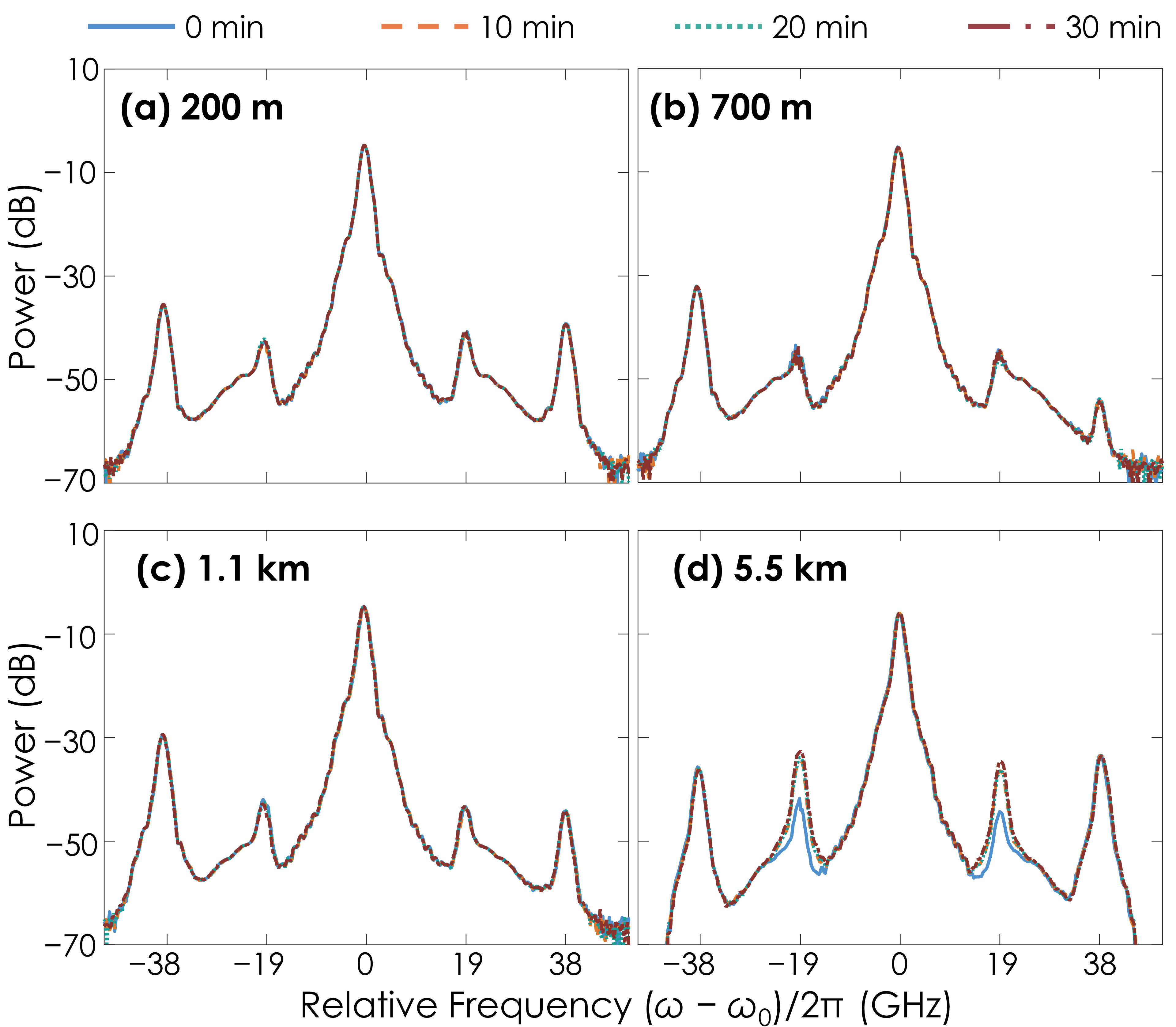}
    \caption{Measured optical spectra for the classical modulation cancellation experiments using our home-built RFoF system. Different lengths of fiber spools are inserted between the two EOPMs to test the system's performance.}
    \label{spools}
\end{figure}

\section{Classical Tests with Fiber Spools Inserted}
\label{plots}

Figure~\ref{spools} presents the optical spectra recorded during classical modulation cancellation tests using the home-built RFoF system over a duration of 30 min. Fiber spools of 200~m, 700~m, 1.1~km, and 5.5~km are inserted between the two EOPMs to emulate the performance for truly physically separated nodes. 

Trials with the 200~m, 700~m, and 1.1~km spools exhibited stability comparable to the trial without a spool. However, the $\sim$9 dB sideband drift during the 5.5~km test is clearly visible on the OSA trace in Fig.~\ref{spools}(d). The cause of this drift remains undetermined, but possible factors include polarization drift and significant dispersion in the reference channel due to the long fiber length---both of which could potentially be mitigated with additional optical components.

\section{Conversion between Sideband Suppression and Timing Offsets}
\label{AppSim}

\begin{figure}[t!]  \centering\includegraphics[width=\columnwidth]{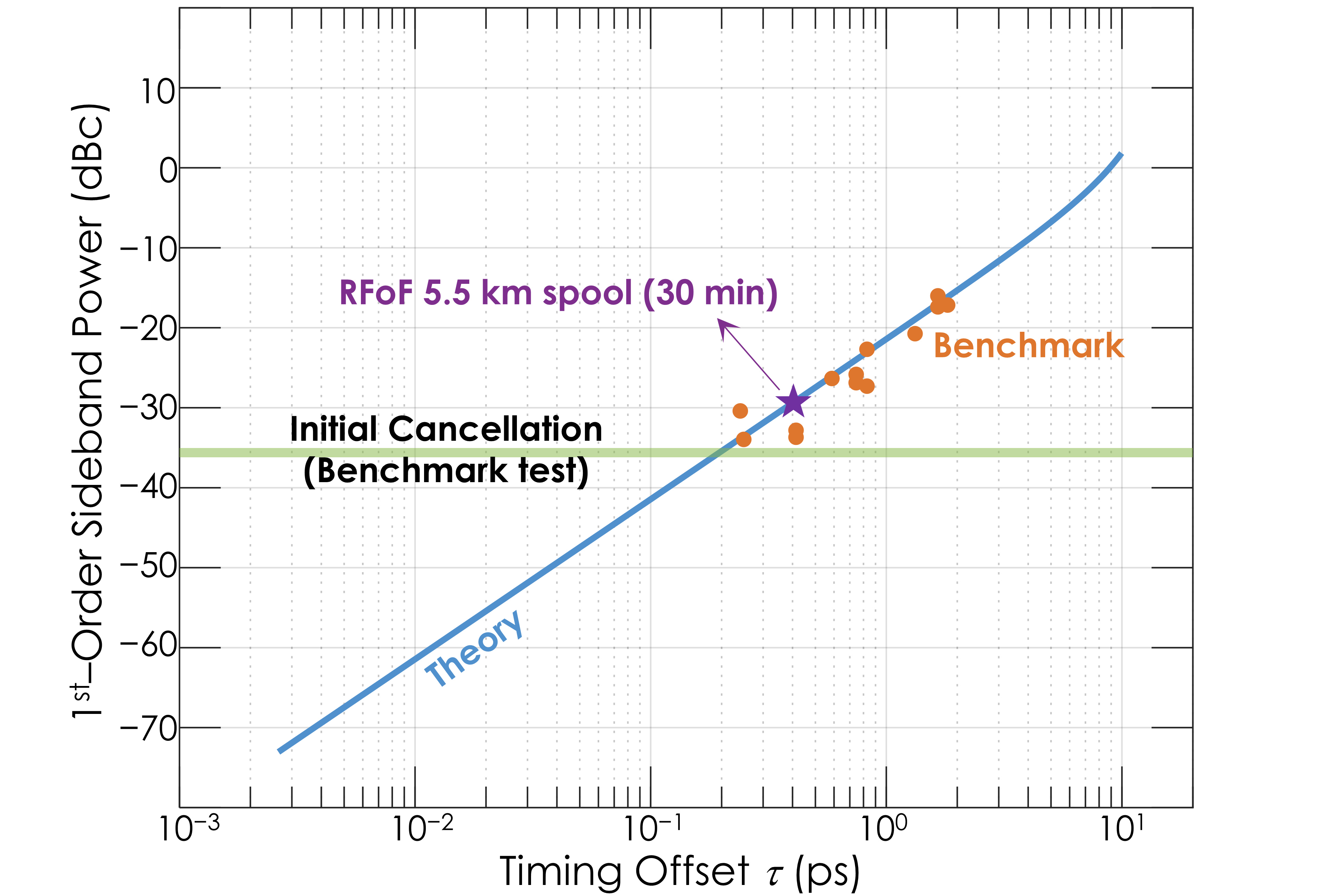}
    \caption{Mapping the contrast of sideband suppression to the timing offset $\tau$ between the two RF signals. Error bars are smaller than marker size. 
    Refer to Appendix~\ref{AppSim} for details.}
    \label{conversion}
\end{figure}

In the classical modulation cancellation experiments, we obtain a series of optical spectra representing different degrees of synchronization between two nodes. Our objective here is to develop a model to infer the amount of timing drift between two RF signals from these spectra. In our simulation, we consider a CW laser (modeled as a delta function in the frequency domain) subjected to two back-to-back temporal phase modulations, both pure 19~GHz sinewaves with a modulation depth of 1.42 rad, matching the settings used in our experiments. We start with the two modulations exactly $180^{\circ}$ out of phase, resulting in perfect cancellation and the original frequency line at the output (i.e., sideband cancellation of $-\infty$ dBc). We then repeat this calculation 500 times, each time increasing the timing offset $\tau$ between them. As the offset increases, the power in the sidebands grows, with the first-order sidebands remaining the most prominent compared to higher orders. The power in the first-order sideband at different values of $\tau$ is depicted by the solid line in Fig.~\ref{conversion}.

For benchmarking, we revisit the experimental setup shown in Fig.~\ref{classic}(a,b), but drive each EOPM by an individual oscillator (Agilent E8257D) synchronized by a shared 10~MHz reference transmitted via a short BNC cable. The oscillator allows us to quantitatively adjust phase with an accuracy of 0.002~rad ($\sim$17~fs at 19~GHz). We fine-tune the phase until maximize the sideband suppression at around $-35$~dBc (green area in Fig.~\ref{conversion}). This performance is potentially limited by several factors, including imperfect dispersion compensation, slight distortion in the RF waveforms, and delay fluctuations in the optical fibers. Comparing this level to the theoretical model implies that this method is capable of resolving timing offsets down to $\sim$0.2~ps, below which other nonidealities dominate the first-order sideband values.
(Additionally, we believe these factors might contribute to the slight asymmetry in the sidebands and the second-order sidebands observed in Figs.~\ref{classic}(c) and \ref{spools}. However, it appears that they do not have any significant impact on the overall performance of the system.)

Once achieving optimal sideband suppression, we immediately adjust the phase setting in the second oscillator's output (to avoid any unintended drifts between two RF signals), and manually introduce a series of $\tau$ values between the two RF signals. For each $\tau$, we record the corresponding optical power in the carrier and the first-order sidebands on the OSA. These data points (shown as orange circles in Fig.~\ref{conversion}) closely match the simulation curve, thereby independently validating the efficacy of our model for retrieving the timing offset from the classical modulation cancellation experiment (purple star in Fig.~\ref{conversion}).

\section{Jitter and the Quantum Frequency Processor}
\label{sec:appC}
In this paper, we have focused on the fundamental quantum mechanical effect of nonlocal electro-optic modulation as representative of distributed quantum processing in the frequency domain. However, as shown through extensive theoretical~\cite{Lukens2017} and experimental~\cite{Lu2018a, Lu2018b, Lu2023c} work on the QFP, EOPMs form just one piece of a larger vision for generic frequency-based quantum information processing. In the QFP paradigm, EOPMs driven periodically at the bin spacing $\Omega$ alternate with line-by-line Fourier transform pulse shapers to synthesize arbitrary quantum unitaries. In the formalism of linear-optical quantum computing~\cite{Knill2001, Kok2007}, the annihilation operators $a_n$ ($b_m$) corresponding to frequency bins centered at $\omega_n=\omega_0+n\Omega$ ($\omega_m=\omega_0+m\Omega$) in the input (output) Hilbert space are related by a mode transformation matrix $V$:
\begin{equation}
\label{eq:J1}
b_m = \sum_{m=-\infty}^{\infty} V_{mn}a_n.
\end{equation}
In the case of a line-by-line pulse shaper, $V_{mn} = e^{i\phi_m}\delta_{mn}$, where $\delta_{mn}$ is the Kronecker delta (equal to unity when $m=n$, zero otherwise). For an EOPM driven by temporal phase $\varphi(t)$, $V_{mn} = c_{m-n}$, where 
\begin{equation}
\label{eq:J2}
c_{m-n} = \frac{1}{T} \int_T dt\, e^{i\varphi(t)} e^{i(m-n)\Omega t}
\end{equation}
corresponds to the Fourier series coefficient of the exponentiated modulation: $e^{i\varphi(t)}=\sum_k c_k e^{-ik\Omega t}$. Here $T=2\pi/\Omega$ denotes the RF period, and integration is performed over any length-$T$ interval.

Concatenation of any set of $Q$ elements follows simply from matrix multiplication, i.e., $W = V^{(Q)} V^{(Q-1)} \cdots V^{(1)}$, where each $V^{(q)}$ corresponds to a specific EOPM or pulse shaper; in practice these infinite-dimensional matrices can be truncated to a dimension sufficiently large to encompass all probability amplitudes of interest~\cite{Lu2023c}.

Now suppose that the RF clock for this QFP is delayed by $\tau$; the time-stationary phases of each pulse shaper remain unaffected, but the EOPM modulation coefficients change to
\begin{align}
\label{eq:J3}
\tilde{c}_{m-n} & = \frac{1}{T} \int_T dt \, e^{i\varphi(t-\tau)} e^{i(m-n)\Omega t} \nonumber\\
 & = \frac{1}{T} \int_T dt \, e^{i\varphi(t^\prime)} e^{i(m-n)\Omega (t^\prime+\tau)} \\
 & = e^{i(m-n)\Omega\tau} c_{m-n}. \nonumber
\end{align}
The matrix multiplication process and the opposite signs on the $m$ and $n$ phase contributions combine to cancel all $\tau$-dependent phases except those on the first and last EOPM; thus the elements of the complete ``$\tau$-shifted'' QFP matrix $\widetilde{W}$ are related to those of $W$ as
\begin{equation}
\label{eq:J4}
\widetilde{W}_{mn} = e^{i(m-n)\Omega\tau} W_{mn}.
\end{equation}
The sensitivity to drift is therefore related not only to the bin spacing $\Omega$, but also the optical bandwidth of interest. That is, for a $d$-dimensional space of frequency bins, $|m-n| < d$, implying $\tau\ll (d\Omega)^{-1}$ as a sufficient condition to maintain high fidelity with respect to the ideal $\tau=0$ case.

Equation~\eqref{eq:J4} applies generically to any QFP operation and can be used to estimate the feasibility of distributed frequency-bin processing under any practical synchronization constraints. But the \emph{quantitative} impact depends on both the specific operation $W$ and selected application. For example, following the QFP immediately with frequency-resolved detection---which is insensitive to the output phase $e^{im\Omega\tau}$---will impose less stringent requirements on jitter than, e.g., a QFP operating as a relay between two other nodes such that both input and output phases matter.

Notwithstanding such application-specific questions, valuable insights can be gained through a concrete example. Consider $W$ corresponding to an ideal $d$-point DFT with components $W_{mn} = \frac{1}{\sqrt{d}} e^{2\pi i (m-n)/d}$ ($m,n\in\{0,...,d-1\})$. The DFT not only proves an excellent test case due to its uniform mixing weight across the entire $d$-bin space; it also admits efficient synthesis via RF harmonic addition in a fixed three-element QFP, with $d\in\{2,3\}$ demonstrated experimentally and $d\leq 10$ designed in simulation~\cite{Lu2018a, Lu2022a}. Considering the matrix fidelity~\cite{Uskov2009,Lukens2017, Lu2023c} between the ideal $W$ and time-shifted $\widetilde{W}$, we find:
\begin{equation}
\label{eq:J5}
\mathcal{F} = \frac{\left| \Tr \widetilde{W}^\dagger W \right|^2}{\Tr \widetilde{W}^\dagger \widetilde{W} \Tr W^\dagger W} =\frac{1}{d^4} \frac{\sin^4 \frac{d\Omega\tau}{2} }{\sin^4 \frac{\Omega\tau}{2} }.
\end{equation}

\begin{figure}[t!]
\vspace{0.2in}
    \includegraphics[width=\columnwidth]{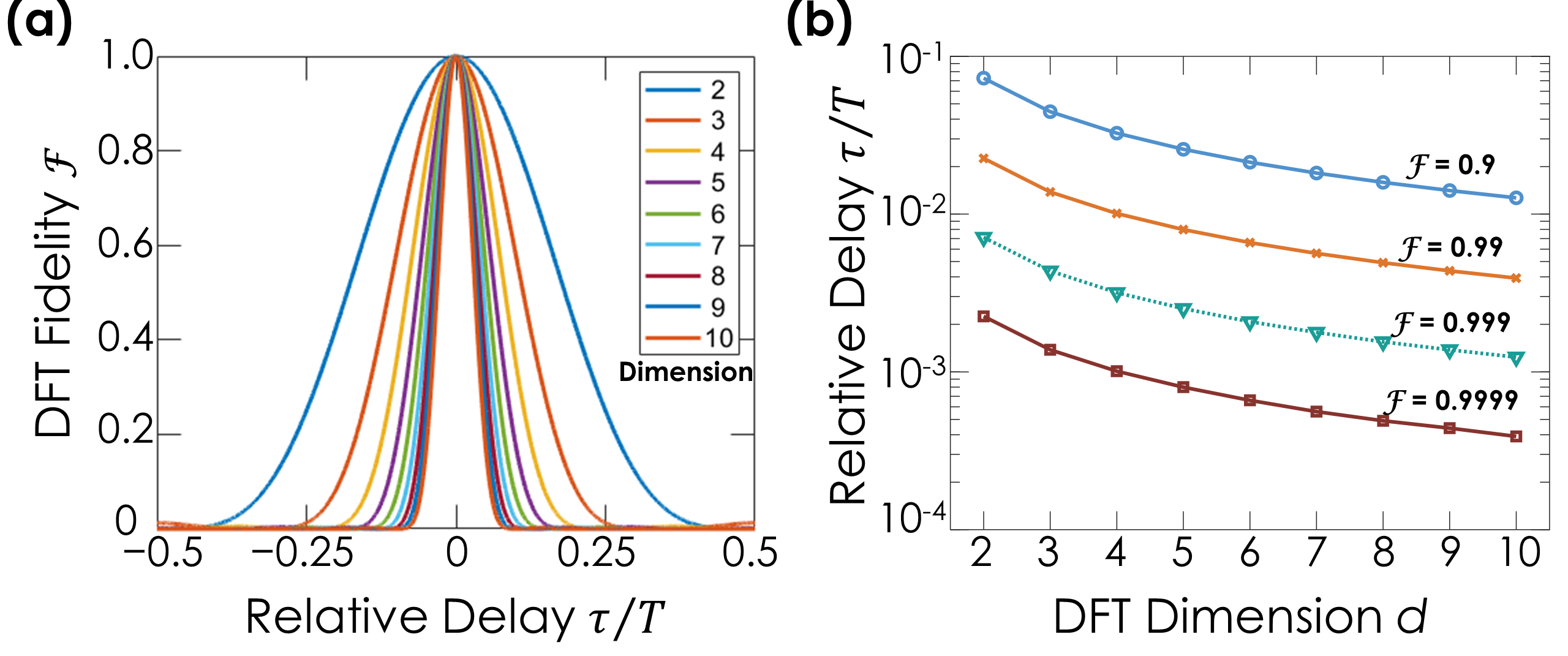}
    \caption{Impact of additional RF delay $\tau$ on QFP performance. (a) Fidelity of simulated $d$-dimensional DFT gates. (b) Maximum tolerable RF delay for various fidelity thresholds.}
    \label{FidelitySims}
\end{figure}

Figure~\ref{FidelitySims} shows the scaling of $\mathcal{F}$ with $\tau$ for $d\in\{2,...,10\}$. As expected from Eq.~\eqref{eq:J4}, sensitivity to $\tau$ becomes increasingly pronounced as the bandwidth (namely, dimension $d$) grows. Incidentally, our experimentally determined bound of $\tau<0.5$~ps for a 5.5~km spool corresponds to $\Omega\tau < 0.02\pi$ (i.e., $\tau/T < 0.01$) at 19~GHz---or $\mathcal{F}>0.9980$ for $d=2$ and $\mathcal{F}>0.9367$ for $d=10$. Consequently, our system in its current form should support high-fidelity, high-dimensional frequency-bin operations over metro-area distances.

Finally, although for simplicity we have focused on ``instantanenous'' fidelity in these calculations---i.e., for a fixed but nonzero $\tau$---one can readily generalize to any temporal distribution through the theory of nonunitary quantum processes. For example, a fluctuating delay with distribution $f(\tau)$ will transform an input qudit density matrix $\rho$ to the output $\sigma$ according to the quantum channel
\begin{equation}
\label{eq:J6}
\sigma = \mathcal{E}(\rho) =\int d\tau\, f(\tau) \widetilde{W}(\tau) \rho \widetilde{W}^\dagger(\tau),
\end{equation}
which models such a drifting QFP in a quantum mechanically complete way~\cite{Lu2023c}.

\providecommand{\noopsort}[1]{}\providecommand{\singleletter}[1]{#1}%


\begin{thebibliography}{56}%
\makeatletter
\providecommand \@ifxundefined [1]{%
 \@ifx{#1\undefined}
}%
\providecommand \@ifnum [1]{%
 \ifnum #1\expandafter \@firstoftwo
 \else \expandafter \@secondoftwo
 \fi
}%
\providecommand \@ifx [1]{%
 \ifx #1\expandafter \@firstoftwo
 \else \expandafter \@secondoftwo
 \fi
}%
\providecommand \natexlab [1]{#1}%
\providecommand \enquote  [1]{``#1''}%
\providecommand \bibnamefont  [1]{#1}%
\providecommand \bibfnamefont [1]{#1}%
\providecommand \citenamefont [1]{#1}%
\providecommand \href@noop [0]{\@secondoftwo}%
\providecommand \href [0]{\begingroup \@sanitize@url \@href}%
\providecommand \@href[1]{\@@startlink{#1}\@@href}%
\providecommand \@@href[1]{\endgroup#1\@@endlink}%
\providecommand \@sanitize@url [0]{\catcode `\\12\catcode `\$12\catcode `\&12\catcode `\#12\catcode `\^12\catcode `\_12\catcode `\%12\relax}%
\providecommand \@@startlink[1]{}%
\providecommand \@@endlink[0]{}%
\providecommand \url  [0]{\begingroup\@sanitize@url \@url }%
\providecommand \@url [1]{\endgroup\@href {#1}{\urlprefix }}%
\providecommand \urlprefix  [0]{URL }%
\providecommand \Eprint [0]{\href }%
\providecommand \doibase [0]{https://doi.org/}%
\providecommand \selectlanguage [0]{\@gobble}%
\providecommand \bibinfo  [0]{\@secondoftwo}%
\providecommand \bibfield  [0]{\@secondoftwo}%
\providecommand \translation [1]{[#1]}%
\providecommand \BibitemOpen [0]{}%
\providecommand \bibitemStop [0]{}%
\providecommand \bibitemNoStop [0]{.\EOS\space}%
\providecommand \EOS [0]{\spacefactor3000\relax}%
\providecommand \BibitemShut  [1]{\csname bibitem#1\endcsname}%
\let\auto@bib@innerbib\@empty
\bibitem [{\citenamefont {Einstein}\ \emph {et~al.}(1935)\citenamefont {Einstein}, \citenamefont {Podolsky},\ and\ \citenamefont {Rosen}}]{Einstein1935}%
  \BibitemOpen
  \bibfield  {author} {\bibinfo {author} {\bibfnamefont {A.}~\bibnamefont {Einstein}}, \bibinfo {author} {\bibfnamefont {B.}~\bibnamefont {Podolsky}},\ and\ \bibinfo {author} {\bibfnamefont {N.}~\bibnamefont {Rosen}},\ }\bibfield  {title} {\bibinfo {title} {Can quantum-mechanical description of physical reality be considered complete?},\ }\href {https://doi.org/10.1103/PhysRev.47.777} {\bibfield  {journal} {\bibinfo  {journal} {Phys. Rev.}\ }\textbf {\bibinfo {volume} {47}},\ \bibinfo {pages} {777} (\bibinfo {year} {1935})}\BibitemShut {NoStop}%
\bibitem [{\citenamefont {Bell}(1964)}]{Bell1964}%
  \BibitemOpen
  \bibfield  {author} {\bibinfo {author} {\bibfnamefont {J.~S.}\ \bibnamefont {Bell}},\ }\bibfield  {title} {\bibinfo {title} {On the {E}instein {P}oldolsky {R}osen {P}aradox},\ }\href@noop {} {\bibfield  {journal} {\bibinfo  {journal} {Physics (Long Island City, NY)}\ }\textbf {\bibinfo {volume} {1}},\ \bibinfo {pages} {195} (\bibinfo {year} {1964})}\BibitemShut {NoStop}%
\bibitem [{\citenamefont {Maudlin}(2019)}]{Maudlin2019}%
  \BibitemOpen
  \bibfield  {author} {\bibinfo {author} {\bibfnamefont {T.}~\bibnamefont {Maudlin}},\ }\href@noop {} {\emph {\bibinfo {title} {Philosophy of Physics: Quantum Theory}}}\ (\bibinfo  {publisher} {Princeton University Press},\ \bibinfo {year} {2019})\BibitemShut {NoStop}%
\bibitem [{\citenamefont {Aspect}\ \emph {et~al.}(1981)\citenamefont {Aspect}, \citenamefont {Grangier},\ and\ \citenamefont {Roger}}]{Aspect1981}%
  \BibitemOpen
  \bibfield  {author} {\bibinfo {author} {\bibfnamefont {A.}~\bibnamefont {Aspect}}, \bibinfo {author} {\bibfnamefont {P.}~\bibnamefont {Grangier}},\ and\ \bibinfo {author} {\bibfnamefont {G.}~\bibnamefont {Roger}},\ }\bibfield  {title} {\bibinfo {title} {Experimental tests of realistic local theories via {Bell's} theorem},\ }\href {https://doi.org/10.1103/PhysRevLett.47.460} {\bibfield  {journal} {\bibinfo  {journal} {Phys. Rev. Lett.}\ }\textbf {\bibinfo {volume} {47}},\ \bibinfo {pages} {460} (\bibinfo {year} {1981})}\BibitemShut {NoStop}%
\bibitem [{\citenamefont {Aspect}\ \emph {et~al.}(1982{\natexlab{a}})\citenamefont {Aspect}, \citenamefont {Grangier},\ and\ \citenamefont {Roger}}]{Aspect1982a}%
  \BibitemOpen
  \bibfield  {author} {\bibinfo {author} {\bibfnamefont {A.}~\bibnamefont {Aspect}}, \bibinfo {author} {\bibfnamefont {P.}~\bibnamefont {Grangier}},\ and\ \bibinfo {author} {\bibfnamefont {G.}~\bibnamefont {Roger}},\ }\bibfield  {title} {\bibinfo {title} {Experimental realization of {Einstein-Podolsky-Rosen-Bohm} gedankenexperiment: A new violation of {Bell's} inequalities},\ }\href {https://doi.org/10.1103/PhysRevLett.49.91} {\bibfield  {journal} {\bibinfo  {journal} {Phys. Rev. Lett.}\ }\textbf {\bibinfo {volume} {49}},\ \bibinfo {pages} {91} (\bibinfo {year} {1982}{\natexlab{a}})}\BibitemShut {NoStop}%
\bibitem [{\citenamefont {Aspect}\ \emph {et~al.}(1982{\natexlab{b}})\citenamefont {Aspect}, \citenamefont {Dalibard},\ and\ \citenamefont {Roger}}]{Aspect1982b}%
  \BibitemOpen
  \bibfield  {author} {\bibinfo {author} {\bibfnamefont {A.}~\bibnamefont {Aspect}}, \bibinfo {author} {\bibfnamefont {J.}~\bibnamefont {Dalibard}},\ and\ \bibinfo {author} {\bibfnamefont {G.}~\bibnamefont {Roger}},\ }\bibfield  {title} {\bibinfo {title} {Experimental test of {Bell's} inequalities using time-varying analyzers},\ }\href {https://doi.org/10.1103/PhysRevLett.49.1804} {\bibfield  {journal} {\bibinfo  {journal} {Phys. Rev. Lett.}\ }\textbf {\bibinfo {volume} {49}},\ \bibinfo {pages} {1804} (\bibinfo {year} {1982}{\natexlab{b}})}\BibitemShut {NoStop}%
\bibitem [{\citenamefont {Hensen}\ \emph {et~al.}(2015)\citenamefont {Hensen}, \citenamefont {Bernien}, \citenamefont {Dreau}, \citenamefont {Reiserer}, \citenamefont {Kalb}, \citenamefont {Blok}, \citenamefont {Ruitenberg}, \citenamefont {Vermeulen}, \citenamefont {Schouten}, \citenamefont {Abellan}, \citenamefont {Amaya}, \citenamefont {Pruneri}, \citenamefont {Mitchell}, \citenamefont {Markham}, \citenamefont {Twitchen}, \citenamefont {Elkouss}, \citenamefont {Wehner}, \citenamefont {Taminiau},\ and\ \citenamefont {Hanson}}]{Hensen2015}%
  \BibitemOpen
  \bibfield  {author} {\bibinfo {author} {\bibfnamefont {B.}~\bibnamefont {Hensen}}, \bibinfo {author} {\bibfnamefont {H.}~\bibnamefont {Bernien}}, \bibinfo {author} {\bibfnamefont {A.~E.}\ \bibnamefont {Dreau}}, \bibinfo {author} {\bibfnamefont {A.}~\bibnamefont {Reiserer}}, \bibinfo {author} {\bibfnamefont {N.}~\bibnamefont {Kalb}}, \bibinfo {author} {\bibfnamefont {M.~S.}\ \bibnamefont {Blok}}, \bibinfo {author} {\bibfnamefont {J.}~\bibnamefont {Ruitenberg}}, \bibinfo {author} {\bibfnamefont {R.~F.~L.}\ \bibnamefont {Vermeulen}}, \bibinfo {author} {\bibfnamefont {R.~N.}\ \bibnamefont {Schouten}}, \bibinfo {author} {\bibfnamefont {C.}~\bibnamefont {Abellan}}, \bibinfo {author} {\bibfnamefont {W.}~\bibnamefont {Amaya}}, \bibinfo {author} {\bibfnamefont {V.}~\bibnamefont {Pruneri}}, \bibinfo {author} {\bibfnamefont {M.~W.}\ \bibnamefont {Mitchell}}, \bibinfo {author} {\bibfnamefont {M.}~\bibnamefont {Markham}}, \bibinfo {author} {\bibfnamefont {D.~J.}\ \bibnamefont {Twitchen}}, \bibinfo {author}
  {\bibfnamefont {D.}~\bibnamefont {Elkouss}}, \bibinfo {author} {\bibfnamefont {S.}~\bibnamefont {Wehner}}, \bibinfo {author} {\bibfnamefont {T.~H.}\ \bibnamefont {Taminiau}},\ and\ \bibinfo {author} {\bibfnamefont {R.}~\bibnamefont {Hanson}},\ }\bibfield  {title} {\bibinfo {title} {Loophole-free {Bell} inequality violation using electron spins separated by 1.3 kilometres},\ }\href {https://doi.org/10.1038/nature15759} {\bibfield  {journal} {\bibinfo  {journal} {Nature}\ }\textbf {\bibinfo {volume} {526}},\ \bibinfo {pages} {682} (\bibinfo {year} {2015})}\BibitemShut {NoStop}%
\bibitem [{\citenamefont {Giustina}\ \emph {et~al.}(2015)\citenamefont {Giustina}, \citenamefont {Versteegh}, \citenamefont {Wengerowsky}, \citenamefont {Handsteiner}, \citenamefont {Hochrainer}, \citenamefont {Phelan}, \citenamefont {Steinlechner}, \citenamefont {Kofler}, \citenamefont {Larsson}, \citenamefont {Abell\'an}, \citenamefont {Amaya}, \citenamefont {Pruneri}, \citenamefont {Mitchell}, \citenamefont {Beyer}, \citenamefont {Gerrits}, \citenamefont {Lita}, \citenamefont {Shalm}, \citenamefont {Nam}, \citenamefont {Scheidl}, \citenamefont {Ursin}, \citenamefont {Wittmann},\ and\ \citenamefont {Zeilinger}}]{Giustina2015}%
  \BibitemOpen
  \bibfield  {author} {\bibinfo {author} {\bibfnamefont {M.}~\bibnamefont {Giustina}}, \bibinfo {author} {\bibfnamefont {M.~A.~M.}\ \bibnamefont {Versteegh}}, \bibinfo {author} {\bibfnamefont {S.}~\bibnamefont {Wengerowsky}}, \bibinfo {author} {\bibfnamefont {J.}~\bibnamefont {Handsteiner}}, \bibinfo {author} {\bibfnamefont {A.}~\bibnamefont {Hochrainer}}, \bibinfo {author} {\bibfnamefont {K.}~\bibnamefont {Phelan}}, \bibinfo {author} {\bibfnamefont {F.}~\bibnamefont {Steinlechner}}, \bibinfo {author} {\bibfnamefont {J.}~\bibnamefont {Kofler}}, \bibinfo {author} {\bibfnamefont {J.-A.}\ \bibnamefont {Larsson}}, \bibinfo {author} {\bibfnamefont {C.}~\bibnamefont {Abell\'an}}, \bibinfo {author} {\bibfnamefont {W.}~\bibnamefont {Amaya}}, \bibinfo {author} {\bibfnamefont {V.}~\bibnamefont {Pruneri}}, \bibinfo {author} {\bibfnamefont {M.~W.}\ \bibnamefont {Mitchell}}, \bibinfo {author} {\bibfnamefont {J.}~\bibnamefont {Beyer}}, \bibinfo {author} {\bibfnamefont {T.}~\bibnamefont {Gerrits}}, \bibinfo {author}
  {\bibfnamefont {A.~E.}\ \bibnamefont {Lita}}, \bibinfo {author} {\bibfnamefont {L.~K.}\ \bibnamefont {Shalm}}, \bibinfo {author} {\bibfnamefont {S.~W.}\ \bibnamefont {Nam}}, \bibinfo {author} {\bibfnamefont {T.}~\bibnamefont {Scheidl}}, \bibinfo {author} {\bibfnamefont {R.}~\bibnamefont {Ursin}}, \bibinfo {author} {\bibfnamefont {B.}~\bibnamefont {Wittmann}},\ and\ \bibinfo {author} {\bibfnamefont {A.}~\bibnamefont {Zeilinger}},\ }\bibfield  {title} {\bibinfo {title} {Significant-loophole-free test of {Bell's} theorem with entangled photons},\ }\href {https://doi.org/10.1103/PhysRevLett.115.250401} {\bibfield  {journal} {\bibinfo  {journal} {Phys. Rev. Lett.}\ }\textbf {\bibinfo {volume} {115}},\ \bibinfo {pages} {250401} (\bibinfo {year} {2015})}\BibitemShut {NoStop}%
\bibitem [{\citenamefont {Shalm}\ \emph {et~al.}(2015)\citenamefont {Shalm}, \citenamefont {Meyer-Scott}, \citenamefont {Christensen}, \citenamefont {Bierhorst}, \citenamefont {Wayne}, \citenamefont {Stevens}, \citenamefont {Gerrits}, \citenamefont {Glancy}, \citenamefont {Hamel}, \citenamefont {Allman}, \citenamefont {Coakley}, \citenamefont {Dyer}, \citenamefont {Hodge}, \citenamefont {Lita}, \citenamefont {Verma}, \citenamefont {Lambrocco}, \citenamefont {Tortorici}, \citenamefont {Migdall}, \citenamefont {Zhang}, \citenamefont {Kumor}, \citenamefont {Farr}, \citenamefont {Marsili}, \citenamefont {Shaw}, \citenamefont {Stern}, \citenamefont {Abell\'an}, \citenamefont {Amaya}, \citenamefont {Pruneri}, \citenamefont {Jennewein}, \citenamefont {Mitchell}, \citenamefont {Kwiat}, \citenamefont {Bienfang}, \citenamefont {Mirin}, \citenamefont {Knill},\ and\ \citenamefont {Nam}}]{Shalm2015}%
  \BibitemOpen
  \bibfield  {author} {\bibinfo {author} {\bibfnamefont {L.~K.}\ \bibnamefont {Shalm}}, \bibinfo {author} {\bibfnamefont {E.}~\bibnamefont {Meyer-Scott}}, \bibinfo {author} {\bibfnamefont {B.~G.}\ \bibnamefont {Christensen}}, \bibinfo {author} {\bibfnamefont {P.}~\bibnamefont {Bierhorst}}, \bibinfo {author} {\bibfnamefont {M.~A.}\ \bibnamefont {Wayne}}, \bibinfo {author} {\bibfnamefont {M.~J.}\ \bibnamefont {Stevens}}, \bibinfo {author} {\bibfnamefont {T.}~\bibnamefont {Gerrits}}, \bibinfo {author} {\bibfnamefont {S.}~\bibnamefont {Glancy}}, \bibinfo {author} {\bibfnamefont {D.~R.}\ \bibnamefont {Hamel}}, \bibinfo {author} {\bibfnamefont {M.~S.}\ \bibnamefont {Allman}}, \bibinfo {author} {\bibfnamefont {K.~J.}\ \bibnamefont {Coakley}}, \bibinfo {author} {\bibfnamefont {S.~D.}\ \bibnamefont {Dyer}}, \bibinfo {author} {\bibfnamefont {C.}~\bibnamefont {Hodge}}, \bibinfo {author} {\bibfnamefont {A.~E.}\ \bibnamefont {Lita}}, \bibinfo {author} {\bibfnamefont {V.~B.}\ \bibnamefont {Verma}}, \bibinfo {author}
  {\bibfnamefont {C.}~\bibnamefont {Lambrocco}}, \bibinfo {author} {\bibfnamefont {E.}~\bibnamefont {Tortorici}}, \bibinfo {author} {\bibfnamefont {A.~L.}\ \bibnamefont {Migdall}}, \bibinfo {author} {\bibfnamefont {Y.}~\bibnamefont {Zhang}}, \bibinfo {author} {\bibfnamefont {D.~R.}\ \bibnamefont {Kumor}}, \bibinfo {author} {\bibfnamefont {W.~H.}\ \bibnamefont {Farr}}, \bibinfo {author} {\bibfnamefont {F.}~\bibnamefont {Marsili}}, \bibinfo {author} {\bibfnamefont {M.~D.}\ \bibnamefont {Shaw}}, \bibinfo {author} {\bibfnamefont {J.~A.}\ \bibnamefont {Stern}}, \bibinfo {author} {\bibfnamefont {C.}~\bibnamefont {Abell\'an}}, \bibinfo {author} {\bibfnamefont {W.}~\bibnamefont {Amaya}}, \bibinfo {author} {\bibfnamefont {V.}~\bibnamefont {Pruneri}}, \bibinfo {author} {\bibfnamefont {T.}~\bibnamefont {Jennewein}}, \bibinfo {author} {\bibfnamefont {M.~W.}\ \bibnamefont {Mitchell}}, \bibinfo {author} {\bibfnamefont {P.~G.}\ \bibnamefont {Kwiat}}, \bibinfo {author} {\bibfnamefont {J.~C.}\ \bibnamefont {Bienfang}},
  \bibinfo {author} {\bibfnamefont {R.~P.}\ \bibnamefont {Mirin}}, \bibinfo {author} {\bibfnamefont {E.}~\bibnamefont {Knill}},\ and\ \bibinfo {author} {\bibfnamefont {S.~W.}\ \bibnamefont {Nam}},\ }\bibfield  {title} {\bibinfo {title} {Strong loophole-free test of local realism},\ }\href {https://doi.org/10.1103/PhysRevLett.115.250402} {\bibfield  {journal} {\bibinfo  {journal} {Phys. Rev. Lett.}\ }\textbf {\bibinfo {volume} {115}},\ \bibinfo {pages} {250402} (\bibinfo {year} {2015})}\BibitemShut {NoStop}%
\bibitem [{\citenamefont {Ekert}(1991)}]{Ekert1991}%
  \BibitemOpen
  \bibfield  {author} {\bibinfo {author} {\bibfnamefont {A.~K.}\ \bibnamefont {Ekert}},\ }\bibfield  {title} {\bibinfo {title} {Quantum cryptography based on {B}ell's theorem},\ }\href {https://doi.org/10.1103/PhysRevLett.67.661} {\bibfield  {journal} {\bibinfo  {journal} {Phys. Rev. Lett.}\ }\textbf {\bibinfo {volume} {67}},\ \bibinfo {pages} {661} (\bibinfo {year} {1991})}\BibitemShut {NoStop}%
\bibitem [{\citenamefont {Gisin}\ \emph {et~al.}(2002)\citenamefont {Gisin}, \citenamefont {Ribordy}, \citenamefont {Tittel},\ and\ \citenamefont {Zbinden}}]{Gisin2002}%
  \BibitemOpen
  \bibfield  {author} {\bibinfo {author} {\bibfnamefont {N.}~\bibnamefont {Gisin}}, \bibinfo {author} {\bibfnamefont {G.}~\bibnamefont {Ribordy}}, \bibinfo {author} {\bibfnamefont {W.}~\bibnamefont {Tittel}},\ and\ \bibinfo {author} {\bibfnamefont {H.}~\bibnamefont {Zbinden}},\ }\bibfield  {title} {\bibinfo {title} {Quantum cryptography},\ }\href {https://doi.org/10.1103/RevModPhys.74.145} {\bibfield  {journal} {\bibinfo  {journal} {Rev. Mod. Phys.}\ }\textbf {\bibinfo {volume} {74}},\ \bibinfo {pages} {145} (\bibinfo {year} {2002})}\BibitemShut {NoStop}%
\bibitem [{\citenamefont {Bennett}\ \emph {et~al.}(1993)\citenamefont {Bennett}, \citenamefont {Brassard}, \citenamefont {Cr\'epeau}, \citenamefont {Jozsa}, \citenamefont {Peres},\ and\ \citenamefont {Wootters}}]{Bennett1993}%
  \BibitemOpen
  \bibfield  {author} {\bibinfo {author} {\bibfnamefont {C.~H.}\ \bibnamefont {Bennett}}, \bibinfo {author} {\bibfnamefont {G.}~\bibnamefont {Brassard}}, \bibinfo {author} {\bibfnamefont {C.}~\bibnamefont {Cr\'epeau}}, \bibinfo {author} {\bibfnamefont {R.}~\bibnamefont {Jozsa}}, \bibinfo {author} {\bibfnamefont {A.}~\bibnamefont {Peres}},\ and\ \bibinfo {author} {\bibfnamefont {W.~K.}\ \bibnamefont {Wootters}},\ }\bibfield  {title} {\bibinfo {title} {Teleporting an unknown quantum state via dual classical and {E}instein-{P}odolsky-{R}osen channels},\ }\href {https://doi.org/10.1103/PhysRevLett.70.1895} {\bibfield  {journal} {\bibinfo  {journal} {Phys. Rev. Lett.}\ }\textbf {\bibinfo {volume} {70}},\ \bibinfo {pages} {1895} (\bibinfo {year} {1993})}\BibitemShut {NoStop}%
\bibitem [{\citenamefont {Bouwmeester}\ \emph {et~al.}(1997)\citenamefont {Bouwmeester}, \citenamefont {Pan}, \citenamefont {Mattle}, \citenamefont {Eibl}, \citenamefont {Weinfurter},\ and\ \citenamefont {Zeilinger}}]{Bouwmeester1997}%
  \BibitemOpen
  \bibfield  {author} {\bibinfo {author} {\bibfnamefont {D.}~\bibnamefont {Bouwmeester}}, \bibinfo {author} {\bibfnamefont {J.-W.}\ \bibnamefont {Pan}}, \bibinfo {author} {\bibfnamefont {K.}~\bibnamefont {Mattle}}, \bibinfo {author} {\bibfnamefont {M.}~\bibnamefont {Eibl}}, \bibinfo {author} {\bibfnamefont {H.}~\bibnamefont {Weinfurter}},\ and\ \bibinfo {author} {\bibfnamefont {A.}~\bibnamefont {Zeilinger}},\ }\bibfield  {title} {\bibinfo {title} {Experimental quantum teleportation},\ }\href {https://doi.org/10.1038/37539} {\bibfield  {journal} {\bibinfo  {journal} {Nature}\ }\textbf {\bibinfo {volume} {390}},\ \bibinfo {pages} {575} (\bibinfo {year} {1997})}\BibitemShut {NoStop}%
\bibitem [{\citenamefont {Hadfield}(2009)}]{Hadfield2009}%
  \BibitemOpen
  \bibfield  {author} {\bibinfo {author} {\bibfnamefont {R.~H.}\ \bibnamefont {Hadfield}},\ }\bibfield  {title} {\bibinfo {title} {Single-photon detectors for optical quantum information applications},\ }\href {https://doi.org/10.1038/nphoton.2009.230} {\bibfield  {journal} {\bibinfo  {journal} {Nat. Photon.}\ }\textbf {\bibinfo {volume} {3}},\ \bibinfo {pages} {696} (\bibinfo {year} {2009})}\BibitemShut {NoStop}%
\bibitem [{\citenamefont {Korzh}\ \emph {et~al.}(2020)\citenamefont {Korzh}, \citenamefont {Zhao}, \citenamefont {Allmaras}, \citenamefont {Frasca}, \citenamefont {Autry}, \citenamefont {Bersin}, \citenamefont {Beyer}, \citenamefont {Briggs}, \citenamefont {Bumble}, \citenamefont {Colangelo}, \citenamefont {Crouch}, \citenamefont {Dane}, \citenamefont {Gerrits}, \citenamefont {Lita}, \citenamefont {Marsili}, \citenamefont {Moody}, \citenamefont {Peña}, \citenamefont {Ramirez}, \citenamefont {Rezac}, \citenamefont {Sinclair}, \citenamefont {Stevens}, \citenamefont {Velasco}, \citenamefont {Verma}, \citenamefont {Wollman}, \citenamefont {Xie}, \citenamefont {Zhu}, \citenamefont {Hale}, \citenamefont {Spiropulu}, \citenamefont {Silverman}, \citenamefont {Mirin}, \citenamefont {Nam}, \citenamefont {Kozorezov}, \citenamefont {Shaw},\ and\ \citenamefont {Berggren}}]{Korzh2020}%
  \BibitemOpen
  \bibfield  {author} {\bibinfo {author} {\bibfnamefont {B.}~\bibnamefont {Korzh}}, \bibinfo {author} {\bibfnamefont {Q.-Y.}\ \bibnamefont {Zhao}}, \bibinfo {author} {\bibfnamefont {J.~P.}\ \bibnamefont {Allmaras}}, \bibinfo {author} {\bibfnamefont {S.}~\bibnamefont {Frasca}}, \bibinfo {author} {\bibfnamefont {T.~M.}\ \bibnamefont {Autry}}, \bibinfo {author} {\bibfnamefont {E.~A.}\ \bibnamefont {Bersin}}, \bibinfo {author} {\bibfnamefont {A.~D.}\ \bibnamefont {Beyer}}, \bibinfo {author} {\bibfnamefont {R.~M.}\ \bibnamefont {Briggs}}, \bibinfo {author} {\bibfnamefont {B.}~\bibnamefont {Bumble}}, \bibinfo {author} {\bibfnamefont {M.}~\bibnamefont {Colangelo}}, \bibinfo {author} {\bibfnamefont {G.~M.}\ \bibnamefont {Crouch}}, \bibinfo {author} {\bibfnamefont {A.~E.}\ \bibnamefont {Dane}}, \bibinfo {author} {\bibfnamefont {T.}~\bibnamefont {Gerrits}}, \bibinfo {author} {\bibfnamefont {A.~E.}\ \bibnamefont {Lita}}, \bibinfo {author} {\bibfnamefont {F.}~\bibnamefont {Marsili}}, \bibinfo {author} {\bibfnamefont
  {G.}~\bibnamefont {Moody}}, \bibinfo {author} {\bibfnamefont {C.}~\bibnamefont {Peña}}, \bibinfo {author} {\bibfnamefont {E.}~\bibnamefont {Ramirez}}, \bibinfo {author} {\bibfnamefont {J.~D.}\ \bibnamefont {Rezac}}, \bibinfo {author} {\bibfnamefont {N.}~\bibnamefont {Sinclair}}, \bibinfo {author} {\bibfnamefont {M.~J.}\ \bibnamefont {Stevens}}, \bibinfo {author} {\bibfnamefont {A.~E.}\ \bibnamefont {Velasco}}, \bibinfo {author} {\bibfnamefont {V.~B.}\ \bibnamefont {Verma}}, \bibinfo {author} {\bibfnamefont {E.~E.}\ \bibnamefont {Wollman}}, \bibinfo {author} {\bibfnamefont {S.}~\bibnamefont {Xie}}, \bibinfo {author} {\bibfnamefont {D.}~\bibnamefont {Zhu}}, \bibinfo {author} {\bibfnamefont {P.~D.}\ \bibnamefont {Hale}}, \bibinfo {author} {\bibfnamefont {M.}~\bibnamefont {Spiropulu}}, \bibinfo {author} {\bibfnamefont {K.~L.}\ \bibnamefont {Silverman}}, \bibinfo {author} {\bibfnamefont {R.~P.}\ \bibnamefont {Mirin}}, \bibinfo {author} {\bibfnamefont {S.~W.}\ \bibnamefont {Nam}}, \bibinfo {author}
  {\bibfnamefont {A.~G.}\ \bibnamefont {Kozorezov}}, \bibinfo {author} {\bibfnamefont {M.~D.}\ \bibnamefont {Shaw}},\ and\ \bibinfo {author} {\bibfnamefont {K.~K.}\ \bibnamefont {Berggren}},\ }\bibfield  {title} {\bibinfo {title} {Demonstration of sub-3 ps temporal resolution with a superconducting nanowire single-photon detector},\ }\href {https://doi.org/10.1038/s41566-020-0589-x} {\bibfield  {journal} {\bibinfo  {journal} {Nat. Photon.}\ }\textbf {\bibinfo {volume} {14}},\ \bibinfo {pages} {250} (\bibinfo {year} {2020})}\BibitemShut {NoStop}%
\bibitem [{\citenamefont {Islam}\ \emph {et~al.}(2017{\natexlab{a}})\citenamefont {Islam}, \citenamefont {Cahall}, \citenamefont {Aragoneses}, \citenamefont {Lezama}, \citenamefont {Kim},\ and\ \citenamefont {Gauthier}}]{Islam2017a}%
  \BibitemOpen
  \bibfield  {author} {\bibinfo {author} {\bibfnamefont {N.~T.}\ \bibnamefont {Islam}}, \bibinfo {author} {\bibfnamefont {C.}~\bibnamefont {Cahall}}, \bibinfo {author} {\bibfnamefont {A.}~\bibnamefont {Aragoneses}}, \bibinfo {author} {\bibfnamefont {A.}~\bibnamefont {Lezama}}, \bibinfo {author} {\bibfnamefont {J.}~\bibnamefont {Kim}},\ and\ \bibinfo {author} {\bibfnamefont {D.~J.}\ \bibnamefont {Gauthier}},\ }\bibfield  {title} {\bibinfo {title} {Robust and stable delay interferometers with application to $d$-dimensional time-frequency quantum key distribution},\ }\href {https://doi.org/10.1103/PhysRevApplied.7.044010} {\bibfield  {journal} {\bibinfo  {journal} {Phys. Rev. Applied}\ }\textbf {\bibinfo {volume} {7}},\ \bibinfo {pages} {044010} (\bibinfo {year} {2017}{\natexlab{a}})}\BibitemShut {NoStop}%
\bibitem [{\citenamefont {Islam}\ \emph {et~al.}(2017{\natexlab{b}})\citenamefont {Islam}, \citenamefont {Lim}, \citenamefont {Cahall}, \citenamefont {Kim},\ and\ \citenamefont {Gauthier}}]{Islam2017b}%
  \BibitemOpen
  \bibfield  {author} {\bibinfo {author} {\bibfnamefont {N.~T.}\ \bibnamefont {Islam}}, \bibinfo {author} {\bibfnamefont {C.~C.~W.}\ \bibnamefont {Lim}}, \bibinfo {author} {\bibfnamefont {C.}~\bibnamefont {Cahall}}, \bibinfo {author} {\bibfnamefont {J.}~\bibnamefont {Kim}},\ and\ \bibinfo {author} {\bibfnamefont {D.~J.}\ \bibnamefont {Gauthier}},\ }\bibfield  {title} {\bibinfo {title} {Provably secure and high-rate quantum key distribution with time-bin qudits},\ }\href {https://doi.org/10.1126/sciadv.1701491} {\bibfield  {journal} {\bibinfo  {journal} {Sci. Adv.}\ }\textbf {\bibinfo {volume} {3}},\ \bibinfo {pages} {e1701491} (\bibinfo {year} {2017}{\natexlab{b}})}\BibitemShut {NoStop}%
\bibitem [{\citenamefont {Wang}\ \emph {et~al.}(2015)\citenamefont {Wang}, \citenamefont {Yin}, \citenamefont {Chen}, \citenamefont {He}, \citenamefont {Song}, \citenamefont {Li}, \citenamefont {Zhang}, \citenamefont {Zhou}, \citenamefont {Guo},\ and\ \citenamefont {Han}}]{Wang2015}%
  \BibitemOpen
  \bibfield  {author} {\bibinfo {author} {\bibfnamefont {S.}~\bibnamefont {Wang}}, \bibinfo {author} {\bibfnamefont {Z.-Q.}\ \bibnamefont {Yin}}, \bibinfo {author} {\bibfnamefont {W.}~\bibnamefont {Chen}}, \bibinfo {author} {\bibfnamefont {D.-Y.}\ \bibnamefont {He}}, \bibinfo {author} {\bibfnamefont {X.-T.}\ \bibnamefont {Song}}, \bibinfo {author} {\bibfnamefont {H.-W.}\ \bibnamefont {Li}}, \bibinfo {author} {\bibfnamefont {L.-J.}\ \bibnamefont {Zhang}}, \bibinfo {author} {\bibfnamefont {Z.}~\bibnamefont {Zhou}}, \bibinfo {author} {\bibfnamefont {G.-C.}\ \bibnamefont {Guo}},\ and\ \bibinfo {author} {\bibfnamefont {Z.-F.}\ \bibnamefont {Han}},\ }\bibfield  {title} {\bibinfo {title} {Experimental demonstration of a quantum key distribution without signal disturbance monitoring},\ }\href {https://doi.org/10.1038/nphoton.2015.209} {\bibfield  {journal} {\bibinfo  {journal} {Nat. Photon.}\ }\textbf {\bibinfo {volume} {9}},\ \bibinfo {pages} {832} (\bibinfo {year} {2015})}\BibitemShut {NoStop}%
\bibitem [{\citenamefont {Wang}\ \emph {et~al.}(2018)\citenamefont {Wang}, \citenamefont {Yin}, \citenamefont {Chau}, \citenamefont {Chen}, \citenamefont {Wang}, \citenamefont {Guo},\ and\ \citenamefont {Han}}]{Wang2018}%
  \BibitemOpen
  \bibfield  {author} {\bibinfo {author} {\bibfnamefont {S.}~\bibnamefont {Wang}}, \bibinfo {author} {\bibfnamefont {Z.-Q.}\ \bibnamefont {Yin}}, \bibinfo {author} {\bibfnamefont {H.~F.}\ \bibnamefont {Chau}}, \bibinfo {author} {\bibfnamefont {W.}~\bibnamefont {Chen}}, \bibinfo {author} {\bibfnamefont {C.}~\bibnamefont {Wang}}, \bibinfo {author} {\bibfnamefont {G.-C.}\ \bibnamefont {Guo}},\ and\ \bibinfo {author} {\bibfnamefont {Z.-F.}\ \bibnamefont {Han}},\ }\bibfield  {title} {\bibinfo {title} {Proof-of-principle experimental realization of a qubit-like qudit-based quantum key distribution scheme},\ }\href {https://doi.org/10.1088/2058-9565/aaace4} {\bibfield  {journal} {\bibinfo  {journal} {Quantum Sci. Technol.}\ }\textbf {\bibinfo {volume} {3}},\ \bibinfo {pages} {025006} (\bibinfo {year} {2018})}\BibitemShut {NoStop}%
\bibitem [{\citenamefont {Lu}\ \emph {et~al.}(2023{\natexlab{a}})\citenamefont {Lu}, \citenamefont {Liscidini}, \citenamefont {Gaeta}, \citenamefont {Weiner},\ and\ \citenamefont {Lukens}}]{Lu2023c}%
  \BibitemOpen
  \bibfield  {author} {\bibinfo {author} {\bibfnamefont {H.-H.}\ \bibnamefont {Lu}}, \bibinfo {author} {\bibfnamefont {M.}~\bibnamefont {Liscidini}}, \bibinfo {author} {\bibfnamefont {A.~L.}\ \bibnamefont {Gaeta}}, \bibinfo {author} {\bibfnamefont {A.~M.}\ \bibnamefont {Weiner}},\ and\ \bibinfo {author} {\bibfnamefont {J.~M.}\ \bibnamefont {Lukens}},\ }\bibfield  {title} {\bibinfo {title} {Frequency-bin photonic quantum information},\ }\href {https://doi.org/10.1364/optica.506096} {\bibfield  {journal} {\bibinfo  {journal} {Optica}\ }\textbf {\bibinfo {volume} {10}},\ \bibinfo {pages} {1655} (\bibinfo {year} {2023}{\natexlab{a}})}\BibitemShut {NoStop}%
\bibitem [{\citenamefont {Lukens}\ and\ \citenamefont {Lougovski}(2017)}]{Lukens2017}%
  \BibitemOpen
  \bibfield  {author} {\bibinfo {author} {\bibfnamefont {J.~M.}\ \bibnamefont {Lukens}}\ and\ \bibinfo {author} {\bibfnamefont {P.}~\bibnamefont {Lougovski}},\ }\bibfield  {title} {\bibinfo {title} {Frequency-encoded photonic qubits for scalable quantum information processing},\ }\href {https://doi.org/10.1364/OPTICA.4.000008} {\bibfield  {journal} {\bibinfo  {journal} {Optica}\ }\textbf {\bibinfo {volume} {4}},\ \bibinfo {pages} {8} (\bibinfo {year} {2017})}\BibitemShut {NoStop}%
\bibitem [{\citenamefont {Kues}\ \emph {et~al.}(2017)\citenamefont {Kues}, \citenamefont {Reimer}, \citenamefont {Roztocki}, \citenamefont {Cort\'{e}s}, \citenamefont {Sciara}, \citenamefont {Wetzel}, \citenamefont {Zhang}, \citenamefont {Cino}, \citenamefont {Chu}, \citenamefont {Little}, \citenamefont {Moss}, \citenamefont {Caspani}, \citenamefont {Aza\~{n}a},\ and\ \citenamefont {Morandotti}}]{Kues2017}%
  \BibitemOpen
  \bibfield  {author} {\bibinfo {author} {\bibfnamefont {M.}~\bibnamefont {Kues}}, \bibinfo {author} {\bibfnamefont {C.}~\bibnamefont {Reimer}}, \bibinfo {author} {\bibfnamefont {P.}~\bibnamefont {Roztocki}}, \bibinfo {author} {\bibfnamefont {L.~R.}\ \bibnamefont {Cort\'{e}s}}, \bibinfo {author} {\bibfnamefont {S.}~\bibnamefont {Sciara}}, \bibinfo {author} {\bibfnamefont {B.}~\bibnamefont {Wetzel}}, \bibinfo {author} {\bibfnamefont {Y.}~\bibnamefont {Zhang}}, \bibinfo {author} {\bibfnamefont {A.}~\bibnamefont {Cino}}, \bibinfo {author} {\bibfnamefont {S.~T.}\ \bibnamefont {Chu}}, \bibinfo {author} {\bibfnamefont {B.~E.}\ \bibnamefont {Little}}, \bibinfo {author} {\bibfnamefont {D.~J.}\ \bibnamefont {Moss}}, \bibinfo {author} {\bibfnamefont {L.}~\bibnamefont {Caspani}}, \bibinfo {author} {\bibfnamefont {J.}~\bibnamefont {Aza\~{n}a}},\ and\ \bibinfo {author} {\bibfnamefont {R.}~\bibnamefont {Morandotti}},\ }\bibfield  {title} {\bibinfo {title} {On-chip generation of high-dimensional entangled quantum states and
  their coherent control},\ }\href {http://dx.doi.org/10.1038/nature22986} {\bibfield  {journal} {\bibinfo  {journal} {Nature}\ }\textbf {\bibinfo {volume} {546}},\ \bibinfo {pages} {622} (\bibinfo {year} {2017})}\BibitemShut {NoStop}%
\bibitem [{\citenamefont {Lu}\ \emph {et~al.}(2018{\natexlab{a}})\citenamefont {Lu}, \citenamefont {Lukens}, \citenamefont {Peters}, \citenamefont {Odele}, \citenamefont {Leaird}, \citenamefont {Weiner},\ and\ \citenamefont {Lougovski}}]{Lu2018a}%
  \BibitemOpen
  \bibfield  {author} {\bibinfo {author} {\bibfnamefont {H.-H.}\ \bibnamefont {Lu}}, \bibinfo {author} {\bibfnamefont {J.~M.}\ \bibnamefont {Lukens}}, \bibinfo {author} {\bibfnamefont {N.~A.}\ \bibnamefont {Peters}}, \bibinfo {author} {\bibfnamefont {O.~D.}\ \bibnamefont {Odele}}, \bibinfo {author} {\bibfnamefont {D.~E.}\ \bibnamefont {Leaird}}, \bibinfo {author} {\bibfnamefont {A.~M.}\ \bibnamefont {Weiner}},\ and\ \bibinfo {author} {\bibfnamefont {P.}~\bibnamefont {Lougovski}},\ }\bibfield  {title} {\bibinfo {title} {Electro-optic frequency beam splitters and tritters for high-fidelity photonic quantum information processing},\ }\href {https://doi.org/10.1103/PhysRevLett.120.030502} {\bibfield  {journal} {\bibinfo  {journal} {Phys. Rev. Lett.}\ }\textbf {\bibinfo {volume} {120}},\ \bibinfo {pages} {030502} (\bibinfo {year} {2018}{\natexlab{a}})}\BibitemShut {NoStop}%
\bibitem [{\citenamefont {Imany}\ \emph {et~al.}(2018)\citenamefont {Imany}, \citenamefont {Jaramillo-Villegas}, \citenamefont {Odele}, \citenamefont {Han}, \citenamefont {Leaird}, \citenamefont {Lukens}, \citenamefont {Lougovski}, \citenamefont {Qi},\ and\ \citenamefont {Weiner}}]{Imany2018}%
  \BibitemOpen
  \bibfield  {author} {\bibinfo {author} {\bibfnamefont {P.}~\bibnamefont {Imany}}, \bibinfo {author} {\bibfnamefont {J.~A.}\ \bibnamefont {Jaramillo-Villegas}}, \bibinfo {author} {\bibfnamefont {O.~D.}\ \bibnamefont {Odele}}, \bibinfo {author} {\bibfnamefont {K.}~\bibnamefont {Han}}, \bibinfo {author} {\bibfnamefont {D.~E.}\ \bibnamefont {Leaird}}, \bibinfo {author} {\bibfnamefont {J.~M.}\ \bibnamefont {Lukens}}, \bibinfo {author} {\bibfnamefont {P.}~\bibnamefont {Lougovski}}, \bibinfo {author} {\bibfnamefont {M.}~\bibnamefont {Qi}},\ and\ \bibinfo {author} {\bibfnamefont {A.~M.}\ \bibnamefont {Weiner}},\ }\bibfield  {title} {\bibinfo {title} {50-{GHz}-spaced comb of high-dimensional frequency-bin entangled photons from an on-chip silicon nitride microresonator},\ }\href {https://doi.org/10.1364/OE.26.001825} {\bibfield  {journal} {\bibinfo  {journal} {Opt. Express}\ }\textbf {\bibinfo {volume} {26}},\ \bibinfo {pages} {1825} (\bibinfo {year} {2018})}\BibitemShut {NoStop}%
\bibitem [{\citenamefont {Lu}\ \emph {et~al.}(2018{\natexlab{b}})\citenamefont {Lu}, \citenamefont {Lukens}, \citenamefont {Peters}, \citenamefont {Williams}, \citenamefont {Weiner},\ and\ \citenamefont {Lougovski}}]{Lu2018b}%
  \BibitemOpen
  \bibfield  {author} {\bibinfo {author} {\bibfnamefont {H.-H.}\ \bibnamefont {Lu}}, \bibinfo {author} {\bibfnamefont {J.~M.}\ \bibnamefont {Lukens}}, \bibinfo {author} {\bibfnamefont {N.~A.}\ \bibnamefont {Peters}}, \bibinfo {author} {\bibfnamefont {B.~P.}\ \bibnamefont {Williams}}, \bibinfo {author} {\bibfnamefont {A.~M.}\ \bibnamefont {Weiner}},\ and\ \bibinfo {author} {\bibfnamefont {P.}~\bibnamefont {Lougovski}},\ }\bibfield  {title} {\bibinfo {title} {Quantum interference and correlation control of frequency-bin qubits},\ }\href {https://doi.org/10.1364/OPTICA.5.001455} {\bibfield  {journal} {\bibinfo  {journal} {Optica}\ }\textbf {\bibinfo {volume} {5}},\ \bibinfo {pages} {1455} (\bibinfo {year} {2018}{\natexlab{b}})}\BibitemShut {NoStop}%
\bibitem [{\citenamefont {Harris}(2008)}]{Harris2008}%
  \BibitemOpen
  \bibfield  {author} {\bibinfo {author} {\bibfnamefont {S.~E.}\ \bibnamefont {Harris}},\ }\bibfield  {title} {\bibinfo {title} {Nonlocal modulation of entangled photons},\ }\href {https://doi.org/10.1103/PhysRevA.78.021807} {\bibfield  {journal} {\bibinfo  {journal} {Phys. Rev. A}\ }\textbf {\bibinfo {volume} {78}},\ \bibinfo {pages} {021807} (\bibinfo {year} {2008})}\BibitemShut {NoStop}%
\bibitem [{Note1()}]{Note1}%
  \BibitemOpen
  \bibinfo {note} {Note that past demonstrations of quantum nonlocal cancellation, including both dispersion~\cite {Franson1992, Baek2009, Lee2014} and modulation cancellation~\cite {Harris2008, Sensarn2009, Zhou2023}, have so far not attempted to close the locality loophole (encountered in Bell tests) by manipulating photons with spacelike separation. Yet nothing fundamentally prevents such spacelike separation from being realized.}\BibitemShut {Stop}%
\bibitem [{\citenamefont {Sensarn}\ \emph {et~al.}(2009)\citenamefont {Sensarn}, \citenamefont {Yin},\ and\ \citenamefont {Harris}}]{Sensarn2009}%
  \BibitemOpen
  \bibfield  {author} {\bibinfo {author} {\bibfnamefont {S.}~\bibnamefont {Sensarn}}, \bibinfo {author} {\bibfnamefont {G.~Y.}\ \bibnamefont {Yin}},\ and\ \bibinfo {author} {\bibfnamefont {S.~E.}\ \bibnamefont {Harris}},\ }\bibfield  {title} {\bibinfo {title} {Observation of nonlocal modulation with entangled photons},\ }\href {https://doi.org/10.1103/PhysRevLett.103.163601} {\bibfield  {journal} {\bibinfo  {journal} {Phys. Rev. Lett.}\ }\textbf {\bibinfo {volume} {103}},\ \bibinfo {pages} {163601} (\bibinfo {year} {2009})}\BibitemShut {NoStop}%
\bibitem [{\citenamefont {Olislager}\ \emph {et~al.}(2010)\citenamefont {Olislager}, \citenamefont {Cussey}, \citenamefont {Nguyen}, \citenamefont {Emplit}, \citenamefont {Massar}, \citenamefont {Merolla},\ and\ \citenamefont {Huy}}]{Olislager2010}%
  \BibitemOpen
  \bibfield  {author} {\bibinfo {author} {\bibfnamefont {L.}~\bibnamefont {Olislager}}, \bibinfo {author} {\bibfnamefont {J.}~\bibnamefont {Cussey}}, \bibinfo {author} {\bibfnamefont {A.~T.}\ \bibnamefont {Nguyen}}, \bibinfo {author} {\bibfnamefont {P.}~\bibnamefont {Emplit}}, \bibinfo {author} {\bibfnamefont {S.}~\bibnamefont {Massar}}, \bibinfo {author} {\bibfnamefont {J.-M.}\ \bibnamefont {Merolla}},\ and\ \bibinfo {author} {\bibfnamefont {K.~P.}\ \bibnamefont {Huy}},\ }\bibfield  {title} {\bibinfo {title} {Frequency-bin entangled photons},\ }\href {https://doi.org/10.1103/PhysRevA.82.013804} {\bibfield  {journal} {\bibinfo  {journal} {Phys. Rev. A}\ }\textbf {\bibinfo {volume} {82}},\ \bibinfo {pages} {013804} (\bibinfo {year} {2010})}\BibitemShut {NoStop}%
\bibitem [{\citenamefont {Sabattoli}\ \emph {et~al.}(2022)\citenamefont {Sabattoli}, \citenamefont {Gianini}, \citenamefont {Simbula}, \citenamefont {Clementi}, \citenamefont {Fincato}, \citenamefont {Boeuf}, \citenamefont {Liscidini}, \citenamefont {Galli},\ and\ \citenamefont {Bajoni}}]{Sabattoli2022}%
  \BibitemOpen
  \bibfield  {author} {\bibinfo {author} {\bibfnamefont {F.~A.}\ \bibnamefont {Sabattoli}}, \bibinfo {author} {\bibfnamefont {L.}~\bibnamefont {Gianini}}, \bibinfo {author} {\bibfnamefont {A.}~\bibnamefont {Simbula}}, \bibinfo {author} {\bibfnamefont {M.}~\bibnamefont {Clementi}}, \bibinfo {author} {\bibfnamefont {A.}~\bibnamefont {Fincato}}, \bibinfo {author} {\bibfnamefont {F.}~\bibnamefont {Boeuf}}, \bibinfo {author} {\bibfnamefont {M.}~\bibnamefont {Liscidini}}, \bibinfo {author} {\bibfnamefont {M.}~\bibnamefont {Galli}},\ and\ \bibinfo {author} {\bibfnamefont {D.}~\bibnamefont {Bajoni}},\ }\bibfield  {title} {\bibinfo {title} {Silicon source of frequency-bin entangled photons},\ }\href {https://doi.org/10.1364/OL.471241} {\bibfield  {journal} {\bibinfo  {journal} {Opt. Lett.}\ }\textbf {\bibinfo {volume} {47}},\ \bibinfo {pages} {6201} (\bibinfo {year} {2022})}\BibitemShut {NoStop}%
\bibitem [{\citenamefont {Seshadri}\ \emph {et~al.}(2022)\citenamefont {Seshadri}, \citenamefont {Lingaraju}, \citenamefont {Lu}, \citenamefont {Imany}, \citenamefont {Leaird},\ and\ \citenamefont {Weiner}}]{Seshadri2022}%
  \BibitemOpen
  \bibfield  {author} {\bibinfo {author} {\bibfnamefont {S.}~\bibnamefont {Seshadri}}, \bibinfo {author} {\bibfnamefont {N.}~\bibnamefont {Lingaraju}}, \bibinfo {author} {\bibfnamefont {H.-H.}\ \bibnamefont {Lu}}, \bibinfo {author} {\bibfnamefont {P.}~\bibnamefont {Imany}}, \bibinfo {author} {\bibfnamefont {D.~E.}\ \bibnamefont {Leaird}},\ and\ \bibinfo {author} {\bibfnamefont {A.~M.}\ \bibnamefont {Weiner}},\ }\bibfield  {title} {\bibinfo {title} {Nonlocal subpicosecond delay metrology using spectral quantum interference},\ }\href {https://doi.org/10.1364/OPTICA.458565} {\bibfield  {journal} {\bibinfo  {journal} {Optica}\ }\textbf {\bibinfo {volume} {9}},\ \bibinfo {pages} {1339} (\bibinfo {year} {2022})}\BibitemShut {NoStop}%
\bibitem [{\citenamefont {Lu}\ \emph {et~al.}(2023{\natexlab{b}})\citenamefont {Lu}, \citenamefont {Alshowkan}, \citenamefont {Myilswamy}, \citenamefont {Weiner}, \citenamefont {Lukens},\ and\ \citenamefont {Peters}}]{Lu2023b}%
  \BibitemOpen
  \bibfield  {author} {\bibinfo {author} {\bibfnamefont {H.-H.}\ \bibnamefont {Lu}}, \bibinfo {author} {\bibfnamefont {M.}~\bibnamefont {Alshowkan}}, \bibinfo {author} {\bibfnamefont {K.~V.}\ \bibnamefont {Myilswamy}}, \bibinfo {author} {\bibfnamefont {A.~M.}\ \bibnamefont {Weiner}}, \bibinfo {author} {\bibfnamefont {J.~M.}\ \bibnamefont {Lukens}},\ and\ \bibinfo {author} {\bibfnamefont {N.~A.}\ \bibnamefont {Peters}},\ }\bibfield  {title} {\bibinfo {title} {Generation and characterization of ultrabroadband polarization--frequency hyperentangled photons},\ }\href {https://doi.org/10.1364/OL.503127} {\bibfield  {journal} {\bibinfo  {journal} {Opt. Lett.}\ }\textbf {\bibinfo {volume} {48}},\ \bibinfo {pages} {6031} (\bibinfo {year} {2023}{\natexlab{b}})}\BibitemShut {NoStop}%
\bibitem [{\citenamefont {Borghi}\ \emph {et~al.}(2023)\citenamefont {Borghi}, \citenamefont {Tagliavacche}, \citenamefont {Sabattoli}, \citenamefont {Dirani}, \citenamefont {Youssef}, \citenamefont {Petit-Etienne}, \citenamefont {Pargon}, \citenamefont {Sipe}, \citenamefont {Liscidini}, \citenamefont {Sciancalepore}, \citenamefont {Galli},\ and\ \citenamefont {Bajoni}}]{Borghi2023}%
  \BibitemOpen
  \bibfield  {author} {\bibinfo {author} {\bibfnamefont {M.}~\bibnamefont {Borghi}}, \bibinfo {author} {\bibfnamefont {N.}~\bibnamefont {Tagliavacche}}, \bibinfo {author} {\bibfnamefont {F.~A.}\ \bibnamefont {Sabattoli}}, \bibinfo {author} {\bibfnamefont {H.~E.}\ \bibnamefont {Dirani}}, \bibinfo {author} {\bibfnamefont {L.}~\bibnamefont {Youssef}}, \bibinfo {author} {\bibfnamefont {C.}~\bibnamefont {Petit-Etienne}}, \bibinfo {author} {\bibfnamefont {E.}~\bibnamefont {Pargon}}, \bibinfo {author} {\bibfnamefont {J.}~\bibnamefont {Sipe}}, \bibinfo {author} {\bibfnamefont {M.}~\bibnamefont {Liscidini}}, \bibinfo {author} {\bibfnamefont {C.}~\bibnamefont {Sciancalepore}}, \bibinfo {author} {\bibfnamefont {M.}~\bibnamefont {Galli}},\ and\ \bibinfo {author} {\bibfnamefont {D.}~\bibnamefont {Bajoni}},\ }\bibfield  {title} {\bibinfo {title} {Reconfigurable silicon photonic chip for the generation of frequency-bin-entangled qudits},\ }\href {https://doi.org/10.1103/PhysRevApplied.19.064026} {\bibfield  {journal}
  {\bibinfo  {journal} {Phys. Rev. Appl.}\ }\textbf {\bibinfo {volume} {19}},\ \bibinfo {pages} {064026} (\bibinfo {year} {2023})}\BibitemShut {NoStop}%
\bibitem [{\citenamefont {Clementi}\ \emph {et~al.}(2023)\citenamefont {Clementi}, \citenamefont {Sabattoli}, \citenamefont {Borghi}, \citenamefont {Gianini}, \citenamefont {Tagliavacche}, \citenamefont {Dirani}, \citenamefont {Youssef}, \citenamefont {Bergamasco}, \citenamefont {Petit-Etienne}, \citenamefont {Pargon}, \citenamefont {Sipe}, \citenamefont {Liscidini}, \citenamefont {Sciancalepore}, \citenamefont {Galli},\ and\ \citenamefont {Bajoni}}]{Clementi2023}%
  \BibitemOpen
  \bibfield  {author} {\bibinfo {author} {\bibfnamefont {M.}~\bibnamefont {Clementi}}, \bibinfo {author} {\bibfnamefont {F.~A.}\ \bibnamefont {Sabattoli}}, \bibinfo {author} {\bibfnamefont {M.}~\bibnamefont {Borghi}}, \bibinfo {author} {\bibfnamefont {L.}~\bibnamefont {Gianini}}, \bibinfo {author} {\bibfnamefont {N.}~\bibnamefont {Tagliavacche}}, \bibinfo {author} {\bibfnamefont {H.~E.}\ \bibnamefont {Dirani}}, \bibinfo {author} {\bibfnamefont {L.}~\bibnamefont {Youssef}}, \bibinfo {author} {\bibfnamefont {N.}~\bibnamefont {Bergamasco}}, \bibinfo {author} {\bibfnamefont {C.}~\bibnamefont {Petit-Etienne}}, \bibinfo {author} {\bibfnamefont {E.}~\bibnamefont {Pargon}}, \bibinfo {author} {\bibfnamefont {J.~E.}\ \bibnamefont {Sipe}}, \bibinfo {author} {\bibfnamefont {M.}~\bibnamefont {Liscidini}}, \bibinfo {author} {\bibfnamefont {C.}~\bibnamefont {Sciancalepore}}, \bibinfo {author} {\bibfnamefont {M.}~\bibnamefont {Galli}},\ and\ \bibinfo {author} {\bibfnamefont {D.}~\bibnamefont {Bajoni}},\ }\bibfield  {title}
  {\bibinfo {title} {Programmable frequency-bin quantum states in a nano-engineered silicon device},\ }\href {https://doi.org/10.1038/s41467-022-35773-6} {\bibfield  {journal} {\bibinfo  {journal} {Nat. Commun.}\ }\textbf {\bibinfo {volume} {14}},\ \bibinfo {pages} {176} (\bibinfo {year} {2023})}\BibitemShut {NoStop}%
\bibitem [{\citenamefont {Yao}(2009)}]{Yao2009}%
  \BibitemOpen
  \bibfield  {author} {\bibinfo {author} {\bibfnamefont {J.}~\bibnamefont {Yao}},\ }\bibfield  {title} {\bibinfo {title} {Microwave photonics},\ }\href {https://doi.org/10.1109/JLT.2008.2009551} {\bibfield  {journal} {\bibinfo  {journal} {J. Light. Technol.}\ }\textbf {\bibinfo {volume} {27}},\ \bibinfo {pages} {314} (\bibinfo {year} {2009})}\BibitemShut {NoStop}%
\bibitem [{\citenamefont {Urick}\ \emph {et~al.}(2015)\citenamefont {Urick}, \citenamefont {McKinney},\ and\ \citenamefont {Williams}}]{Urick2015}%
  \BibitemOpen
  \bibfield  {author} {\bibinfo {author} {\bibfnamefont {V.~J.}\ \bibnamefont {Urick}, \bibfnamefont {Jr.}}, \bibinfo {author} {\bibfnamefont {J.~D.}\ \bibnamefont {McKinney}},\ and\ \bibinfo {author} {\bibfnamefont {K.~J.}\ \bibnamefont {Williams}},\ }\href@noop {} {\emph {\bibinfo {title} {Fundamentals of Microwave Photonics}}}\ (\bibinfo  {publisher} {Wiley},\ \bibinfo {address} {Hoboken, NJ},\ \bibinfo {year} {2015})\BibitemShut {NoStop}%
\bibitem [{\citenamefont {{Lipi\'{n}ski}}\ \emph {et~al.}(2011)\citenamefont {{Lipi\'{n}ski}}, \citenamefont {{W\l ostowski}}, \citenamefont {{Serrano}},\ and\ \citenamefont {{Alvarez}}}]{Lipinski2011}%
  \BibitemOpen
  \bibfield  {author} {\bibinfo {author} {\bibfnamefont {M.}~\bibnamefont {{Lipi\'{n}ski}}}, \bibinfo {author} {\bibfnamefont {T.}~\bibnamefont {{W\l ostowski}}}, \bibinfo {author} {\bibfnamefont {J.}~\bibnamefont {{Serrano}}},\ and\ \bibinfo {author} {\bibfnamefont {P.}~\bibnamefont {{Alvarez}}},\ }\bibfield  {title} {\bibinfo {title} {{White Rabbit}: a {PTP} application for robust sub-nanosecond synchronization},\ }in\ \href {https://doi.org/10.1109/ISPCS.2011.6070148} {\emph {\bibinfo {booktitle} {IEEE Int. Sym. Precision Clock Sync. Meas. Contr. Commun.}}}\ (\bibinfo {year} {2011})\ pp.\ \bibinfo {pages} {25--30}\BibitemShut {NoStop}%
\bibitem [{IEE(2020)}]{IEEE2019}%
  \BibitemOpen
  \bibfield  {title} {\bibinfo {title} {{IEEE} standard for a precision clock synchronization protocol for networked measurement and control systems},\ }\href {https://doi.org/10.1109/IEEESTD.2020.9120376} {\bibfield  {journal} {\bibinfo  {journal} {IEEE Standard}\ }\textbf {\bibinfo {volume} {1588-2019}} (\bibinfo {year} {2020})}\BibitemShut {NoStop}%
\bibitem [{\citenamefont {Rizzi}\ \emph {et~al.}(2018)\citenamefont {Rizzi}, \citenamefont {Lipinski}, \citenamefont {Ferrari}, \citenamefont {Rinaldi},\ and\ \citenamefont {Flammini}}]{Rizzi2018}%
  \BibitemOpen
  \bibfield  {author} {\bibinfo {author} {\bibfnamefont {M.}~\bibnamefont {Rizzi}}, \bibinfo {author} {\bibfnamefont {M.}~\bibnamefont {Lipinski}}, \bibinfo {author} {\bibfnamefont {P.}~\bibnamefont {Ferrari}}, \bibinfo {author} {\bibfnamefont {S.}~\bibnamefont {Rinaldi}},\ and\ \bibinfo {author} {\bibfnamefont {A.}~\bibnamefont {Flammini}},\ }\bibfield  {title} {\bibinfo {title} {{White} {Rabbit} clock synchronization: Ultimate limits on close-in phase noise and short-term stability due to {FPGA} implementation},\ }\href {https://doi.org/10.1109/TUFFC.2018.2851842} {\bibfield  {journal} {\bibinfo  {journal} {IEEE Trans. Ultrason. Ferroelectr. Freq. Control.}\ }\textbf {\bibinfo {volume} {65}},\ \bibinfo {pages} {1726} (\bibinfo {year} {2018})}\BibitemShut {NoStop}%
\bibitem [{\citenamefont {Alshowkan}\ \emph {et~al.}(2022)\citenamefont {Alshowkan}, \citenamefont {Evans}, \citenamefont {Williams}, \citenamefont {Rao}, \citenamefont {Marvinney}, \citenamefont {Pai}, \citenamefont {Lawrie}, \citenamefont {Peters},\ and\ \citenamefont {Lukens}}]{Alshowkan2022b}%
  \BibitemOpen
  \bibfield  {author} {\bibinfo {author} {\bibfnamefont {M.}~\bibnamefont {Alshowkan}}, \bibinfo {author} {\bibfnamefont {P.~G.}\ \bibnamefont {Evans}}, \bibinfo {author} {\bibfnamefont {B.~P.}\ \bibnamefont {Williams}}, \bibinfo {author} {\bibfnamefont {N.~S.~V.}\ \bibnamefont {Rao}}, \bibinfo {author} {\bibfnamefont {C.~E.}\ \bibnamefont {Marvinney}}, \bibinfo {author} {\bibfnamefont {Y.-Y.}\ \bibnamefont {Pai}}, \bibinfo {author} {\bibfnamefont {B.~J.}\ \bibnamefont {Lawrie}}, \bibinfo {author} {\bibfnamefont {N.~A.}\ \bibnamefont {Peters}},\ and\ \bibinfo {author} {\bibfnamefont {J.~M.}\ \bibnamefont {Lukens}},\ }\bibfield  {title} {\bibinfo {title} {Advanced architectures for high-performance quantum networking},\ }\href {https://doi.org/10.1364/JOCN.450201} {\bibfield  {journal} {\bibinfo  {journal} {J. Opt. Commun. Netw.}\ }\textbf {\bibinfo {volume} {14}},\ \bibinfo {pages} {493} (\bibinfo {year} {2022})}\BibitemShut {NoStop}%
\bibitem [{\citenamefont {Burenkov}\ \emph {et~al.}(2023)\citenamefont {Burenkov}, \citenamefont {Semionov}, \citenamefont {Hala}, \citenamefont {Gerrits}, \citenamefont {Rahmouni}, \citenamefont {Anand}, \citenamefont {Li-Baboud}, \citenamefont {Slattery}, \citenamefont {Battou},\ and\ \citenamefont {Polyakov}}]{Burenkov2023}%
  \BibitemOpen
  \bibfield  {author} {\bibinfo {author} {\bibfnamefont {I.~A.}\ \bibnamefont {Burenkov}}, \bibinfo {author} {\bibfnamefont {A.}~\bibnamefont {Semionov}}, \bibinfo {author} {\bibnamefont {Hala}}, \bibinfo {author} {\bibfnamefont {T.}~\bibnamefont {Gerrits}}, \bibinfo {author} {\bibfnamefont {A.}~\bibnamefont {Rahmouni}}, \bibinfo {author} {\bibfnamefont {D.}~\bibnamefont {Anand}}, \bibinfo {author} {\bibfnamefont {Y.-S.}\ \bibnamefont {Li-Baboud}}, \bibinfo {author} {\bibfnamefont {O.}~\bibnamefont {Slattery}}, \bibinfo {author} {\bibfnamefont {A.}~\bibnamefont {Battou}},\ and\ \bibinfo {author} {\bibfnamefont {S.~V.}\ \bibnamefont {Polyakov}},\ }\bibfield  {title} {\bibinfo {title} {Synchronization and coexistence in quantum networks},\ }\href {https://doi.org/10.1364/OE.480486} {\bibfield  {journal} {\bibinfo  {journal} {Opt. Express}\ }\textbf {\bibinfo {volume} {31}},\ \bibinfo {pages} {11431} (\bibinfo {year} {2023})}\BibitemShut {NoStop}%
\bibitem [{\citenamefont {Foreman}\ \emph {et~al.}(2007)\citenamefont {Foreman}, \citenamefont {Holman}, \citenamefont {Hudson}, \citenamefont {Jones},\ and\ \citenamefont {Ye}}]{Foreman2007}%
  \BibitemOpen
  \bibfield  {author} {\bibinfo {author} {\bibfnamefont {S.~M.}\ \bibnamefont {Foreman}}, \bibinfo {author} {\bibfnamefont {K.~W.}\ \bibnamefont {Holman}}, \bibinfo {author} {\bibfnamefont {D.~D.}\ \bibnamefont {Hudson}}, \bibinfo {author} {\bibfnamefont {D.~J.}\ \bibnamefont {Jones}},\ and\ \bibinfo {author} {\bibfnamefont {J.}~\bibnamefont {Ye}},\ }\bibfield  {title} {\bibinfo {title} {Remote transfer of ultrastable frequency references via fiber networks},\ }\href {https://doi.org/10.1063/1.2437069} {\bibfield  {journal} {\bibinfo  {journal} {Rev. Sci. Instrum.}\ }\textbf {\bibinfo {volume} {78}},\ \bibinfo {pages} {021101} (\bibinfo {year} {2007})}\BibitemShut {NoStop}%
\bibitem [{\citenamefont {Xin}\ \emph {et~al.}(2017)\citenamefont {Xin}, \citenamefont {{\c{S}}afak}, \citenamefont {Peng}, \citenamefont {Kalaydzhyan}, \citenamefont {Wang}, \citenamefont {M\"{u}cke},\ and\ \citenamefont {K\"{a}rtner}}]{Xin2017}%
  \BibitemOpen
  \bibfield  {author} {\bibinfo {author} {\bibfnamefont {M.}~\bibnamefont {Xin}}, \bibinfo {author} {\bibfnamefont {K.}~\bibnamefont {{\c{S}}afak}}, \bibinfo {author} {\bibfnamefont {M.~Y.}\ \bibnamefont {Peng}}, \bibinfo {author} {\bibfnamefont {A.}~\bibnamefont {Kalaydzhyan}}, \bibinfo {author} {\bibfnamefont {W.-T.}\ \bibnamefont {Wang}}, \bibinfo {author} {\bibfnamefont {O.~D.}\ \bibnamefont {M\"{u}cke}},\ and\ \bibinfo {author} {\bibfnamefont {F.~X.}\ \bibnamefont {K\"{a}rtner}},\ }\bibfield  {title} {\bibinfo {title} {Attosecond precision multi-kilometer laser-microwave network},\ }\href {https://doi.org/10.1038/lsa.2016.187} {\bibfield  {journal} {\bibinfo  {journal} {Light Sci. Appl.}\ }\textbf {\bibinfo {volume} {6}},\ \bibinfo {pages} {e16187} (\bibinfo {year} {2017})}\BibitemShut {NoStop}%
\bibitem [{\citenamefont {Lu}\ \emph {et~al.}(2020)\citenamefont {Lu}, \citenamefont {Simmerman}, \citenamefont {Lougovski}, \citenamefont {Weiner},\ and\ \citenamefont {Lukens}}]{Lu2020b}%
  \BibitemOpen
  \bibfield  {author} {\bibinfo {author} {\bibfnamefont {H.-H.}\ \bibnamefont {Lu}}, \bibinfo {author} {\bibfnamefont {E.~M.}\ \bibnamefont {Simmerman}}, \bibinfo {author} {\bibfnamefont {P.}~\bibnamefont {Lougovski}}, \bibinfo {author} {\bibfnamefont {A.~M.}\ \bibnamefont {Weiner}},\ and\ \bibinfo {author} {\bibfnamefont {J.~M.}\ \bibnamefont {Lukens}},\ }\bibfield  {title} {\bibinfo {title} {Fully arbitrary control of frequency-bin qubits},\ }\href {https://doi.org/10.1103/PhysRevLett.125.120503} {\bibfield  {journal} {\bibinfo  {journal} {Phys. Rev. Lett.}\ }\textbf {\bibinfo {volume} {125}},\ \bibinfo {pages} {120503} (\bibinfo {year} {2020})}\BibitemShut {NoStop}%
\bibitem [{Note2()}]{Note2}%
  \BibitemOpen
  \bibinfo {note} {In this work, we define the amount of sideband suppression, or contrast, as the difference between the optical power in the original frequency mode and the power in highest first-order sideband.}\BibitemShut {Stop}%
\bibitem [{\citenamefont {{Pasternack}}(2016)}]{Pasternack}%
  \BibitemOpen
  \bibfield  {author} {\bibinfo {author} {\bibnamefont {{Pasternack}}},\ }\href@noop {} {\bibinfo {title} {Rf cable assemblies technical data sheet}},\ \bibinfo {howpublished} {\url{https://www.pasternack.com/images/ProductPDF/PE3C0230.pdf}} (\bibinfo {year} {2016})\BibitemShut {NoStop}%
\bibitem [{\citenamefont {Chapman}\ \emph {et~al.}(2023)\citenamefont {Chapman}, \citenamefont {Miloshevsky}, \citenamefont {Lu}, \citenamefont {Rao}, \citenamefont {Alshowkan},\ and\ \citenamefont {Peters}}]{Chapman2023}%
  \BibitemOpen
  \bibfield  {author} {\bibinfo {author} {\bibfnamefont {J.~C.}\ \bibnamefont {Chapman}}, \bibinfo {author} {\bibfnamefont {A.}~\bibnamefont {Miloshevsky}}, \bibinfo {author} {\bibfnamefont {H.-H.}\ \bibnamefont {Lu}}, \bibinfo {author} {\bibfnamefont {N.}~\bibnamefont {Rao}}, \bibinfo {author} {\bibfnamefont {M.}~\bibnamefont {Alshowkan}},\ and\ \bibinfo {author} {\bibfnamefont {N.~A.}\ \bibnamefont {Peters}},\ }\bibfield  {title} {\bibinfo {title} {Two-mode squeezing over deployed fiber coexisting with conventional communications},\ }\href@noop {} {\bibfield  {journal} {\bibinfo  {journal} {Opt. Express}\ }\textbf {\bibinfo {volume} {31}},\ \bibinfo {pages} {26254} (\bibinfo {year} {2023})}\BibitemShut {NoStop}%
\bibitem [{\citenamefont {Sheridan}\ and\ \citenamefont {Scarani}(2010)}]{Sheridan2010}%
  \BibitemOpen
  \bibfield  {author} {\bibinfo {author} {\bibfnamefont {L.}~\bibnamefont {Sheridan}}\ and\ \bibinfo {author} {\bibfnamefont {V.}~\bibnamefont {Scarani}},\ }\bibfield  {title} {\bibinfo {title} {Security proof for quantum key distribution using qudit systems},\ }\href {https://doi.org/10.1103/PhysRevA.82.030301} {\bibfield  {journal} {\bibinfo  {journal} {Phys. Rev. A}\ }\textbf {\bibinfo {volume} {82}},\ \bibinfo {pages} {030301} (\bibinfo {year} {2010})}\BibitemShut {NoStop}%
\bibitem [{\citenamefont {Knill}\ \emph {et~al.}(2001)\citenamefont {Knill}, \citenamefont {Laflamme},\ and\ \citenamefont {Milburn}}]{Knill2001}%
  \BibitemOpen
  \bibfield  {author} {\bibinfo {author} {\bibfnamefont {E.}~\bibnamefont {Knill}}, \bibinfo {author} {\bibfnamefont {R.}~\bibnamefont {Laflamme}},\ and\ \bibinfo {author} {\bibfnamefont {G.~J.}\ \bibnamefont {Milburn}},\ }\bibfield  {title} {\bibinfo {title} {A scheme for efficient quantum computation with linear optics},\ }\href {https://doi.org/10.1038/35051009} {\bibfield  {journal} {\bibinfo  {journal} {Nature}\ }\textbf {\bibinfo {volume} {409}},\ \bibinfo {pages} {46} (\bibinfo {year} {2001})}\BibitemShut {NoStop}%
\bibitem [{\citenamefont {Kok}\ \emph {et~al.}(2007)\citenamefont {Kok}, \citenamefont {Munro}, \citenamefont {Nemoto}, \citenamefont {Ralph}, \citenamefont {Dowling},\ and\ \citenamefont {Milburn}}]{Kok2007}%
  \BibitemOpen
  \bibfield  {author} {\bibinfo {author} {\bibfnamefont {P.}~\bibnamefont {Kok}}, \bibinfo {author} {\bibfnamefont {W.~J.}\ \bibnamefont {Munro}}, \bibinfo {author} {\bibfnamefont {K.}~\bibnamefont {Nemoto}}, \bibinfo {author} {\bibfnamefont {T.~C.}\ \bibnamefont {Ralph}}, \bibinfo {author} {\bibfnamefont {J.~P.}\ \bibnamefont {Dowling}},\ and\ \bibinfo {author} {\bibfnamefont {G.~J.}\ \bibnamefont {Milburn}},\ }\bibfield  {title} {\bibinfo {title} {Linear optical quantum computing with photonic qubits},\ }\href {https://doi.org/10.1103/RevModPhys.79.135} {\bibfield  {journal} {\bibinfo  {journal} {Rev. Mod. Phys.}\ }\textbf {\bibinfo {volume} {79}},\ \bibinfo {pages} {135} (\bibinfo {year} {2007})}\BibitemShut {NoStop}%
\bibitem [{\citenamefont {Lu}\ \emph {et~al.}(2022)\citenamefont {Lu}, \citenamefont {Lingaraju}, \citenamefont {Leaird}, \citenamefont {Weiner},\ and\ \citenamefont {Lukens}}]{Lu2022a}%
  \BibitemOpen
  \bibfield  {author} {\bibinfo {author} {\bibfnamefont {H.-H.}\ \bibnamefont {Lu}}, \bibinfo {author} {\bibfnamefont {N.~B.}\ \bibnamefont {Lingaraju}}, \bibinfo {author} {\bibfnamefont {D.~E.}\ \bibnamefont {Leaird}}, \bibinfo {author} {\bibfnamefont {A.~M.}\ \bibnamefont {Weiner}},\ and\ \bibinfo {author} {\bibfnamefont {J.~M.}\ \bibnamefont {Lukens}},\ }\bibfield  {title} {\bibinfo {title} {High-dimensional discrete {Fourier} transform gates with a quantum frequency processor},\ }\href {https://doi.org/10.1364/OE.454677} {\bibfield  {journal} {\bibinfo  {journal} {Opt. Express}\ }\textbf {\bibinfo {volume} {30}},\ \bibinfo {pages} {10126} (\bibinfo {year} {2022})}\BibitemShut {NoStop}%
\bibitem [{\citenamefont {Uskov}\ \emph {et~al.}(2009)\citenamefont {Uskov}, \citenamefont {Kaplan}, \citenamefont {Smith}, \citenamefont {Huver},\ and\ \citenamefont {Dowling}}]{Uskov2009}%
  \BibitemOpen
  \bibfield  {author} {\bibinfo {author} {\bibfnamefont {D.~B.}\ \bibnamefont {Uskov}}, \bibinfo {author} {\bibfnamefont {L.}~\bibnamefont {Kaplan}}, \bibinfo {author} {\bibfnamefont {A.~M.}\ \bibnamefont {Smith}}, \bibinfo {author} {\bibfnamefont {S.~D.}\ \bibnamefont {Huver}},\ and\ \bibinfo {author} {\bibfnamefont {J.~P.}\ \bibnamefont {Dowling}},\ }\bibfield  {title} {\bibinfo {title} {Maximal success probabilities of linear-optical quantum gates},\ }\href {https://doi.org/10.1103/PhysRevA.79.042326} {\bibfield  {journal} {\bibinfo  {journal} {Phys. Rev. A}\ }\textbf {\bibinfo {volume} {79}},\ \bibinfo {pages} {042326} (\bibinfo {year} {2009})}\BibitemShut {NoStop}%
\bibitem [{\citenamefont {Franson}(1992)}]{Franson1992}%
  \BibitemOpen
  \bibfield  {author} {\bibinfo {author} {\bibfnamefont {J.~D.}\ \bibnamefont {Franson}},\ }\bibfield  {title} {\bibinfo {title} {Nonlocal cancellation of dispersion},\ }\href {https://doi.org/10.1103/PhysRevA.45.3126} {\bibfield  {journal} {\bibinfo  {journal} {Phys. Rev. A}\ }\textbf {\bibinfo {volume} {45}},\ \bibinfo {pages} {3126} (\bibinfo {year} {1992})}\BibitemShut {NoStop}%
\bibitem [{\citenamefont {Baek}\ \emph {et~al.}(2009)\citenamefont {Baek}, \citenamefont {Cho},\ and\ \citenamefont {Kim}}]{Baek2009}%
  \BibitemOpen
  \bibfield  {author} {\bibinfo {author} {\bibfnamefont {S.-Y.}\ \bibnamefont {Baek}}, \bibinfo {author} {\bibfnamefont {Y.-W.}\ \bibnamefont {Cho}},\ and\ \bibinfo {author} {\bibfnamefont {Y.-H.}\ \bibnamefont {Kim}},\ }\bibfield  {title} {\bibinfo {title} {Nonlocal dispersion cancellation using entangled photons},\ }\href {https://doi.org/10.1364/OE.17.019241} {\bibfield  {journal} {\bibinfo  {journal} {Opt. Express}\ }\textbf {\bibinfo {volume} {17}},\ \bibinfo {pages} {19241} (\bibinfo {year} {2009})}\BibitemShut {NoStop}%
\bibitem [{\citenamefont {Lee}\ \emph {et~al.}(2014)\citenamefont {Lee}, \citenamefont {Zhang}, \citenamefont {Steinbrecher}, \citenamefont {Zhou}, \citenamefont {Mower}, \citenamefont {Zhong}, \citenamefont {Wang}, \citenamefont {Hu}, \citenamefont {Horansky}, \citenamefont {Verma}, \citenamefont {Lita}, \citenamefont {Mirin}, \citenamefont {Marsili}, \citenamefont {Shaw}, \citenamefont {Nam}, \citenamefont {Wornell}, \citenamefont {Wong}, \citenamefont {Shapiro},\ and\ \citenamefont {Englund}}]{Lee2014}%
  \BibitemOpen
  \bibfield  {author} {\bibinfo {author} {\bibfnamefont {C.}~\bibnamefont {Lee}}, \bibinfo {author} {\bibfnamefont {Z.}~\bibnamefont {Zhang}}, \bibinfo {author} {\bibfnamefont {G.~R.}\ \bibnamefont {Steinbrecher}}, \bibinfo {author} {\bibfnamefont {H.}~\bibnamefont {Zhou}}, \bibinfo {author} {\bibfnamefont {J.}~\bibnamefont {Mower}}, \bibinfo {author} {\bibfnamefont {T.}~\bibnamefont {Zhong}}, \bibinfo {author} {\bibfnamefont {L.}~\bibnamefont {Wang}}, \bibinfo {author} {\bibfnamefont {X.}~\bibnamefont {Hu}}, \bibinfo {author} {\bibfnamefont {R.~D.}\ \bibnamefont {Horansky}}, \bibinfo {author} {\bibfnamefont {V.~B.}\ \bibnamefont {Verma}}, \bibinfo {author} {\bibfnamefont {A.~E.}\ \bibnamefont {Lita}}, \bibinfo {author} {\bibfnamefont {R.~P.}\ \bibnamefont {Mirin}}, \bibinfo {author} {\bibfnamefont {F.}~\bibnamefont {Marsili}}, \bibinfo {author} {\bibfnamefont {M.~D.}\ \bibnamefont {Shaw}}, \bibinfo {author} {\bibfnamefont {S.~W.}\ \bibnamefont {Nam}}, \bibinfo {author} {\bibfnamefont {G.~W.}\ \bibnamefont
  {Wornell}}, \bibinfo {author} {\bibfnamefont {F.~N.~C.}\ \bibnamefont {Wong}}, \bibinfo {author} {\bibfnamefont {J.~H.}\ \bibnamefont {Shapiro}},\ and\ \bibinfo {author} {\bibfnamefont {D.}~\bibnamefont {Englund}},\ }\bibfield  {title} {\bibinfo {title} {Entanglement-based quantum communication secured by nonlocal dispersion cancellation},\ }\href {https://doi.org/10.1103/PhysRevA.90.062331} {\bibfield  {journal} {\bibinfo  {journal} {Phys. Rev. A}\ }\textbf {\bibinfo {volume} {90}},\ \bibinfo {pages} {062331} (\bibinfo {year} {2014})}\BibitemShut {NoStop}%
\bibitem [{\citenamefont {Zhou}\ \emph {et~al.}(2023)\citenamefont {Zhou}, \citenamefont {de~Araujo}, \citenamefont {DiMario}, \citenamefont {Anderson}, \citenamefont {Zhao}, \citenamefont {Jones},\ and\ \citenamefont {Lett}}]{Zhou2023}%
  \BibitemOpen
  \bibfield  {author} {\bibinfo {author} {\bibfnamefont {Z.}~\bibnamefont {Zhou}}, \bibinfo {author} {\bibfnamefont {L.~E.}\ \bibnamefont {de~Araujo}}, \bibinfo {author} {\bibfnamefont {M.}~\bibnamefont {DiMario}}, \bibinfo {author} {\bibfnamefont {B.}~\bibnamefont {Anderson}}, \bibinfo {author} {\bibfnamefont {J.}~\bibnamefont {Zhao}}, \bibinfo {author} {\bibfnamefont {K.~M.}\ \bibnamefont {Jones}},\ and\ \bibinfo {author} {\bibfnamefont {P.~D.}\ \bibnamefont {Lett}},\ }\bibfield  {title} {\bibinfo {title} {Nonlocal phase modulation of multimode, continuous-variable twin beams},\ }\href {https://doi.org/10.1364/OPTICAQ.505870} {\bibfield  {journal} {\bibinfo  {journal} {Optica Quantum}\ }\textbf {\bibinfo {volume} {1}},\ \bibinfo {pages} {71} (\bibinfo {year} {2023})}\BibitemShut {NoStop}%
\end{thebibliography}
\end{document}